\documentclass[%
 aps,
 pre,
 amsmath,amssymb,
reprint,%
graphicx,floatfix]{revtex4-1}

\usepackage{graphicx}% Include figure files
\usepackage{dcolumn}% Align table columns on decimal point
\usepackage{bm}% bold math

\usepackage[utf8]{inputenc}
\usepackage[T1]{fontenc}
\usepackage{mathptmx}
\usepackage{etoolbox}
\usepackage{braket}
\usepackage{xcolor}

 \numberwithin{equation}{section}
 \numberwithin{figure}{section}
 \numberwithin{table}{section}

\usepackage[unicode=true,bookmarks=true,bookmarksopen=false,
	bookmarksopenlevel=1,pdfborder={0 0 0},backref=false,
	colorlinks=true,allcolors=magenta,pdfstartview=Fit,
	pdfcenterwindow=true,pdfdisplaydoctitle=true,
	pdfpagelayout=TwoPageRight,linktocpage=true]{hyperref}

\renewcommand{\tensor}[1]{\overline{\overline{#1}}}

\newcommand{\pd}{\partial}

\newcommand{\w}{\omega}
\newcommand{\W}{\Omega}

\renewcommand{\prl}{\parallel}
\newcommand{\heavy}{`}

	{\begin{inparaenum}[(\itshape i\/\upshape )]} {\end{inparaenum}}

\newcommand{\approptoinn}[2]{\mathrel{\vcenter{
  \offinterlineskip\halign{\hfil$##$\cr
    #1\propto\cr\noalign{\kern2pt}#1\sim\cr\noalign{\kern-2pt}}}}}

\makeatletter
\def\@email#1#2{%
 \endgroup
 \patchcmd{\titleblock@produce}
  {\frontmatter@RRAPformat}
  {\frontmatter@RRAPformat{\produce@RRAP{*#1\href{mailto:#2}{#2}}}\frontmatter@RRAPformat}
  {}{}
}%
\makeatother
\begin{document}

\preprint{APS/123-QED}

\title[]{Lower-Hybrid Drift Instabilities in a magnetic nozzle}
\author{Matteo Ripoli}
\altaffiliation{PhD Student, {mripoli@ing.uc3m.es}}

\author{Eduardo Ahedo}
\altaffiliation{Full Professor, eahedo@ing.uc3m.es}

\author{Mario Merino}
\altaffiliation[]{Full Professor, mario.merino@uc3m.es}
\affiliation{ 
Department of Aerospace Engineering, Universidad Carlos III de Madrid, Leganés, Spain%\\This line break forced with \textbackslash\textbackslash
}%

\date{\today}% It is always \today, today,
 
\begin{abstract}
Magnetic nozzles are a key component of electrodeless plasma thrusters, acting as their main acceleration stage.
Non-stationary phenomena common to the entire range of \textbf{$E \times B$} devices, such as oscillations and instabilities, are likely to exist in the magnetic nozzle, according to the mounting experimental evidence. 
These mechanisms could lead to anomalous cross-field transport, either enhancing the plasma plume divergence or favoring electron detachment.
%, necessary for the operation of the device.
In this work we present a local linear analysis of fluid instabilities relevant for said devices, expanding on previous works with the addition of plasma inhomogeneities in the direction parallel to the magnetic field, with a rigorous inclusion of the effects of magnetic curvature, finite Larmor radius and $3$D wave propagation, allowing for a general formulation of drift-driven instabilities in partially magnetized plasmas. Instability conditions are first studied analytically, and then applied to 
%hybrid PIC/fluid/wave
simulation data of a helicon plasma thruster. Finally, the effect of instabilities on wave-driven cross-field electron transport is assessed by means of quasi-linear analysis. This study predicts the onset of essentially-azimuthal instabilities in the $1$ kHz--$1$ MHz range, in qualitative agreement with some of the available experimental data, and highlights the importance of including parallel inhomogeneities in the formulation of the dispersion relation of an \textbf{$E \times B$} plasma, as these gradients may drive instabilities even in the absence of axial propagation. Lastly, quasi-linear analysis suggests that the induced 
cross-field transport acts to smooth out the zeroth-order drifts which cause the plasma to destabilize in the first place.
\end{abstract}

%\mmm{quizás alguna ecuación la quieres poner en widetext}

%\mmm{Other titles?}\\
%\mrip{Linear and Quasilinear Analysis of Fluid Instabilities in a magnetic nozzle,\\
%Lower-Hybrid Drift Instabilities in a magnetic nozzle\\
%Stability Analysis of a magnetic nozzle}

%\mmm{Mario  as last author?}

%----------------------------------------------------------------------------------
%----------------------------------------------------------------------------------
%----------------------------------------------------------------------------------

\maketitle

%\mmm{Revisar que estamos explicando suficientemente bien que hacemos una expansión plana (cartesiana), y que lo usamos para un problema axisymmetric (cilíndrico); idealmente dar también el límite de validez en función de ktheta, kr, r. (entiendo que kr pequeños y ktheta grandes? o ktheta no importa?)}

%\mmm{Igualmente, revisar el uso de la palabra "azimuthal" (direction). En algunos casos puede tener sentido , en otros no}

%----------------------------------------------------------------------------------
%----------------------------------------------------------------------------------
%----------------------------------------------------------------------------------

\section{Introduction}

A variety of plasma thrusters operate as partially magnetized \textbf{$E \times B$} discharges, with electrons closely following magnetic field lines and the heavier ions moving almost freely from magnetic forces effects. Two notable classes of plasma thrusters are the Hall Thruster (HTs) \cite{kauf85, goeb08, ahed11s}
and Electrodeless Plasma Thrusters (EPTs)\cite{nava18a, pack19} that rely on a magnetic nozzle (MN) \cite{meri16g} as their accelerating stage. 
\textbf{$E \times B$} discharges are known to be subject to oscillations, instabilities, and turbulence \cite{kaga20a}, which
under certain circumstances lead to non-classical transport of electrons across magnetic field lines, suggesting the existence of additional mechanisms not captured by usual steady-state electron models.

There is ample literature on the study of oscillations through local linear analysis, all sharing the common intention of finding sound physical principles and criteria behind onset of instability-driven anomalous transport in plasmas through a limited but analytically accessible formulation \cite{hast03, ahed08i, ramo19, ramo21, hara22, boeu23}.
The assumption of locality is satisfied as long as the wavelength of the considered waves is considerably smaller than the shortest local characteristic length at equilibrium.
For a two-species Maxwellian plasma at 
equilibrium, consisting of unmagnetized ions and magnetized electrons, the choice of studying its oscillations by means of either a fluid or a kinetic approach ultimately falls upon the scale of the considered problem. The fluid approach is generally considered valid as long as the perpendicular wavenumber times the equilibrium electron Larmor radius $\rho_{e0}$ is a small number less than $1$; in the parallel direction, the condition $|k_\prl c_{e0}| < |\w_e|$ needs to be respected, with $k_\prl$, $c_{e0}$ and $\w_e$ the parallel wavenumber, the equilibrium thermal electron velocity, and the wave frequency in the electron reference frame.
This second condition implies that particles moving at a thermal velocity parallel to the magnetic field must be slower than the wave;
at the same time, kinetic particle-wave interactions such as Landau damping are neglected.
The advantages of employing a fluid approach over a kinetic one lie in its reduced complexity \cite{ahed23a}, at the price of assuming \textit{a priori} the aforementioned upper limits on the perpendicular and parallel components of the wave vector $\bm k$.

HT plasmas have been thoroughly studied analytically, numerically and experimentally \cite{chou01b, kaga20a, bell21b, poli23, boeu23, bell24}. Oscillations have been found from the kHz to the tens of MHz ranges.
Morozov et al. \cite{moro72b}
employed a two-species fluid  model with cold, inertialess electrons to justify experimentally observed azimuthally rotating structures. That work is one of the first to study the effect of plasma gradients and relative drift between plasma species on the onset of plasma instabilities, which will be referred to as \textit{drift-gradient instabilities}. It was later expanded on by Esipchuk and Tilinin \cite{esip76}
with electron inertia and electromagnetic effects,
by Frias et al. \cite{fria12}
with inertialess electrons with the addition of density and temperature gradients,
by Escobar and Ahedo \cite{esco14} with the inclusion of neutral dynamics and ionization collisions,
and by Smolyakov et al. \cite{smol17}
with the inclusion of electron inertia and off-diagonal parts of the electron stress tensor, all of them sharing the common focus on HT plasmas.
These last two works included the effect of \textit{drift-resistive instabilities} as well, originating from the combination of relative inter-species drift and collisional effects, previously studied by Litvak and Fisch \cite{litv01} for both electrostatic and electromagnetic waves.
Ramos, Bello and Ahedo \cite{ramo21} provided a more general derivation of fluid electrostatic instabilities in \textbf{$E\times B$} plasmas, examining in detail drift-gradient and drift-resisitive instabilities in a variety of frequency and wavelength regimes along with \textit{stream instabilities}, a class of unstable phenomena uniquely driven by relative drift between species, first presented by Bunemann \cite{bune62}.
%\mrip{still missing enrique, deciding where to put it}
Another work relevant to linear stability analysis of fluid electrostatic waves in HTs but from a global perspective is that of Bello and Ahedo \cite{bell21a}, where numerical results of zeroth and first-order fluid models are obtained. Among the main findings of the work, the effect of including temperature perturbations is found to be non-negligible, but only for frequencies in the MHz range and greater.

Among the few works including wave propagation both along and across magnetic field lines, the one of Krall \cite{kral71} from $1971$ stands out. In the context of general \textbf{$E\times B$} discharges, his proposed kinetic model was able to recover in the negligible Larmor radius limit ($k \rho_{e0} \ll 1$) results known from fluid theory plus a stream instability driven by parallel propagation.

In contast, in the case of MNs and EPTs, not much work has been carried out in the analysis of their unsteady behaviour yet. Recent experimental works have indicated the presence of both azimuthal oscillations extending up to the hundreds of kHz \cite{taka22a,madd24a,madd24b} 
and azimuthal-axial oscillations \cite{hepn20b,vinci22e}.
%\mmm{mete a  \cite{olse15} por aqui}
Desjardins and Gilmore \cite{desj16} observe mainly-azimuthal fluctuations in geometrically-comparable linear plasma devices, in the kHz range.
Various candidate frameworks have been proposed to explain these phenomena, from the destabilization of electrostatic lower hybrid waves in the work of Hepner et al. \cite{hepn20b} to the magnetosonic wave in the work of Takahashi et al. \cite{taka22a}, the latter being electromagnetic in nature.  
In the work of Desjardins and Gilmore \cite{desj16}, the oscillations are identified as a mixture of drift-resistive electron drift waves and Kelvin-Helmoltz instabilities.

Moreover, the role of instabilities in MNs is not clear, and the literature presents clashing arguments on the effect and the direction of wave-driven transport. Hepner et al. \cite{hepn20b} propose an outward electron flux, relaxing the density gradient and reducing the device efficiency, while Takahashi et al. \cite{taka22a} suggest an inward particle flux, pushing the plasma towards its axis of symmetry.

The naming conventions for instabilities in partially magnetized plasmas are abundant, and at times conflicting. 
Two main labelling categories can be identified, relevant to two different frequency regimes: the Electron Cyclotron Drift Instability (ECDI) \cite{boeu23, bell24}, relevant to frequencies  comparable to harmonics of the electron gyrofrequency in the electron frame, and the Lower-Hybrid Drift Instabilities (LHDI) \cite{huba76,smol17,hepn20b}, relevant instead to frequencies close to the lower-hybrid frequency. Notable fluid limits of the latter are the drift-gradient Modified Simon-Hoh Instability (MSHI) \cite{smol17} and the Modified Two-Stream Instability (MTSI) \cite{kral71}.

What stands out from the existing body of work on \textbf{$E\times B$} discharges is that many of the identified instabilities are particular limits of a more general dispersion relation.
In fact, it can be stated that a whole family of \textit{fluid} instabilities stems from the presence of a non-zero interspecies drift, be it gradient-driven or otherwise, allowing the presence of `slow' waves with phase velocity smaller than the drift velocity, or, in other terms, with negative Doppler-shifted frequency.
These waves can be described as carrying negative energy \cite{stur60}: 
when coupled with an energy sink---either a positive energy wave or a dissipative process such as inelastic collisions---they can become unstable \cite{hase75}.
 
All of the above is based on a 1D description of equilibrium gradients, as it is the relevant case in HTs, Penning and magnetron discharges.
In this work we derive a comprehensive formulation for electrostatic drift waves, taking into account inertial, gyroviscous, collisional and 2D gradients effects as well as 3D wave propagation, conditions relevant to MNs and EPTs. The combination of all the aforementioned contributions results in a dispersion relation from which novel instability criteria can be obtained, considerably more general than the ones present in the literature.
Most noticeably, the inclusion of parallel gradients of equilibrium plasma quantities other than perpendicular ones allows for an easier destabilization of the plasma, given the difference in thermal velocities between the species.
The derivation is carried out in a fluid framework, assuming cold, unmagnetized ions and warm electrons. We will assume long wavelengths, with the aforementioned limit on the perpendicular wavenumber and  $|k_\prl c_{e0}| < |\w_e|$, and we will perform a cartesian expansion to study an axisymmetric problem.
We further assume our plasma to have isotropic temperature at equilibrium and neglect temperature oscillations. 
Consquently in the following, for brevity, we drop the subscript $0$ on equilibrium magnetic field and temperature and related magnitudes, so that $B \equiv B_0$, $T_{e} \equiv T_{e0}$, etcetera.

We focus on the low-to-mid frequency range $\w_{ci} \ll \w < \w_{ce}$, with $\w_{cs}$ being the equilibrium cyclotron for the $s$-th species ($i=$ ions, $e=$ electrons), and work with power expansions on the small parameter $\epsilon=\rho_{e} / L \ll 1$, with $L$ being the shortest local characteristic length at equilibrium.
An \textit{a priori} choice has to be made on the order of magnitude of the inertial terms with respect to the cyclotron terms, i.e. whether the electron Doppler-shifted frequency $\w_e \equiv \w - \bm k \cdot \bm u_{e0}$ is comparable with $\w_{ce}$ or with $\w_{ce}\ O(\rho_{e} / L)$. We will refer to the former choice as to the `High-Frequency' (HF) regime, while to the latter as the `Low-Frequency' (LF) one, a differentiation similar to that presented in [\onlinecite{ramo21}].
%\eag{this is used by Ramos alreaady}
%\mmm{cite Ramos as Eduardo suggests}
%
%
The obtained dispersion relation is then applied to hybrid PIC/fluid/wave simulations of MNs from Jimenez et al. \cite{jime23a} as input for the equilibrium plasma quantities and gradients, in order to investigate the eventual triggering for instabilities in these devices.
%\mrip{briefly present the findings}
The predicted instabilities propagate predominantly in the azimuthal direction, with frequencies ranging between $1$ kHz and $1$ MHz. The presence of parallel gradients of equilibrium plasma quantities considerably widens the unstable regions of the discharge, allowing for the onset of exponential wave growth even when the conditions for `classical' instabilities coming the literature (such as the MSHI and the MTSI) are not met.
The findings are qualitatively contrasted with available experimental data. 

%\mmm{revisar}
The rest of the paper is structured as follows: in section \ref{sec:2} we will show the employed fluid model and the general derivation procedure of the dispersion relation in the LF regime.
In section \ref{sec:disp_rel_deriv} we will obtain the detailed formulation of the dispersion relation, comparing it with formulations found in literature.
In section \ref{sec:LF} we will show analytical solutions of the LF dispersion relation, expressing general instability criteria for both drift-gradient and drift-dissipative perturbations.
In section \ref{sec:AD} we will specialize the dispersion relation to a simulated MN plasma
using the data from Jimenez et al. \cite{jime23a} as input for the equilibrium plasma quantities and gradients.
In section \ref{sec:ql} we will model the effect of unstable oscillations on cross-field fluxes and velocities through quasi-linear analysis.
Finally, in section \ref{sec:conclusion} we will present a summary and discuss the main findings of this work. 
A preliminary version of this work has been presented as a conference paper in [\onlinecite{ripo24a}], where the iterative procedure needed to derive the HF dispersion relation was discussed as well.

%\mmm{somewhere: say that symbolic software was used to aid in the derivations; the notebook and intstructions to run it can be found as additional materials to the article in the journal page. Can also say that a preliminary version of this work was presented at ref xyz (iepc)}
 
%----------------------------------------------------------------------------------

\section{Fluid model}

\label{sec:2}

This section presents the derivation procedure of the local, linear, electrostatic dispersion relation for an \textbf{$E\times B$} two-fluid plasma composed of cold, unmagnetized ions $i$, and warm, thermally isotropic, magnetized electrons $e$. The model retains perpendicular and parallel gradients, wave propagation in all three directions, gyroviscous tensor terms and collisional phenomena. The model is consistent up to the $O(\epsilon)$ order. 

Our main focus of application is an axisymmetric MN, with the axis of symmetry coinciding with the $z$ axis,
we define the local coordinate system $\{ \bm 1_\prl, \bm 1_\perp, \bm 1_\theta \}$, 
with $\bm 1_\prl = \bm B/B$, $\bm 1_\theta$ perpendicular to the ($z,r$) meridian plane, and $\bm 1_\perp = \bm 1_\theta \times \bm 1_\parallel$.
Due to the zeroth-order axisymmetry of the discharge,  gradients of the zeroth-order quantities are contained in the $(\bm 1_\prl , \bm 1_\perp)$ plane. 

For our analysis to be local we require the wavenumber to satisfy $k L = k \rho_e / \epsilon \gg 1$. We will make use of a cartesian expansion to study an axisymmetric problem, neglecting cylindrical terms. Thus, the azimuthal wavenumber must respect $|r k_\theta| \gg 1$. 
Being our model fluid, we  limit the normalized perpendicular wavenumber to values $k_{\top} \rho_{e}< 1$, with $k_\top=(k_\perp^2+k_\theta^2)^{1/2}$, $\rho_{e} = c_{e} / \w_{ce}$ the electron gyroradius at equilibrium, $\w_{ce}$ the equilibrium electron gyrofrequency, $c_{e}^2 \equiv {T_{e} / m_e}$ the equilibrium electron thermal velocity
%, $T_{e0}$ the (isotropic) equilibrium electron temperature 
 and $m_e$ the electron mass.   
For the same reason, in the parallel direction, where motion of electron particles is essentially the free thermal drift, the wavelength must be larger than the distance covered by a single particle during an oscillation, a condition which can be expressed as $|k_\prl c_{e}| < |\w_e|$. 
%We  introduce the reference small parameter, $\epsilon = \rho_e / L$, with $L$ the smallest reference equilibrium length in our plasma.
 
%----------------------------------------------------------------------------------
%----------------------------------------------------------------------------------

\subsection{General electron equations}

Warm, magnetized electrons are described by their continuity and momentum equations, which read:
\begin{equation}
    \frac{\pd n_e}{\pd t} + \nabla \cdot \left( n_e \bm{u}_e \right) =
    \nu_p n_e,
    \label{0th:cont}
\end{equation}
\begin{align}
    \frac{\partial \bm{u}_e}{\partial t} + \bm{u}_e \cdot \nabla \bm{u}_e =
    - \frac{\nabla \cdot \tensor p_e}{m_e n_e} - \frac{e}{m_e} \left( -\nabla \phi + \bm{u}_e \times \bm{B} \right) - \nu_{e} \bm{u}_e ,\label{0th:el:mom}
\end{align}
where $\nu_p$ 
%\eag{choose between $\nu_p$ (my choice) or $\nu_P$ through the paper}
represents the particle production rate, $\phi$ the electrostatic potential,   $\nu_{e}$ is used to model dissipative forces on the electrons coming from collisional phenomena, and $\tensor p_e$ 
the complete electron pressure tensor including the gyroviscous contribution. The system has to be completed with the energy equation, which we are not going to consider in the perturbation problem since we are neglecting temperature perturbations, i.e., $T_{e1} = 0$.  This assumption is reasonable for low frequency oscillations \cite{bell21a}.
 
Under the assumption of small amplitude waves, each  quantity $Q$ in the equations above is expanded as a zeroth-order, time independent part, plus a first-order contribution, through which we will model any oscillatory phenomena, 
\begin{align}
Q (\bm x, t) = Q_0 (\bm x) + \frac{1}{2} [Q_1 \left( \bm x \right) \exp\left( i \bm k \cdot \bm x - i \w t \right) + CC],
\end{align}
with $\bm x = s_\perp \bm 1_\perp + s_\theta \bm 1_\theta + s_\prl \bm 1_\prl$, being $s_{\perp, \theta, \prl}$ local coordinates about the point of analysis, and with the subscripts $0$ and $1$ referring to equilibrium values and their first-order corrections, respectively.
The nomenclature $CC$ serves as a reminder that complex conjugates need to be added to recover a real quantity; in the following it is omitted for brevity.
Note that solutions with $\w_{r}<0$ are equivalent to solutions with $\w_{r}>0$ but opposite sign of the components of $\bm k$, with $\w_r \equiv \text{Re} \{ \w \}$.
It is easy to check that the first-order terms of the  gradient of $Q$ is composed of two contributions, 
%\eag{$ i \bm k Q_1 + \nabla Q_1$,  (avoid equation and extra definition)}
%$ i \bm k Q_1 + \nabla Q_1$
%
\begin{align}
i \bm k Q_1 + \nabla Q_1, \label{1st:grad}   
\end{align}
with $|\nabla \ln Q_1| = k\ O\left( \epsilon \right)$ by assumption.
%It is important to state here that we will be looking for shapes of $\nabla \ln Q_1$ that \textit{do not} depend directly on $\w$ and are given as a superposition of the equilibrium gradients. Instead, the wavenumber vector $\bm k$ will be directly related to $\w$ through the dispersion relation $\w=\w(\bm k)$.

\subsection{Zeroth-order equilibrium}

In this work, the
zeroth-order equilibrium plasma quantities and gradients are taken from the hybrid (PIC/fluid/wave) simulations of the MN of a helicon plasma thruster, presented in [\onlinecite{jime23a}].
%\mmm{add Pedro to add the bibtex to his dataset, and cite it here too.}.
The full detail of the model, its numerical implementation, and the results can be found in that work and references therein.
These simulations implement essentially the same electron equations as above, except that they drop electron inertia, temperature anisotropies and gyroviscous terms for the electron equilibrium in Eq. \eqref{0th:el:mom}, consider a scalar electron pressure $\tensor p_{e0} =p_{e0} I$ with $p_{e0}=n_{e0} T_{e}$. They also implement an energy equation for the electron temperature $T_{e}$, together with a heat flux closure, which here is not considered as first-order perturbations will be considered isothermal. Ions and neutrals are treated kinetically as macroparticles. 
%Ionization and electron-neutral collisions are considered.
The collisional mechanisms considered are: single and double ionization; elastic electron-neutral and electron-ion collisions; neutral excitation collisions.
These zeroth-order solutions are just used for a quantitative assessment of the first-order perturbation model.

%---------------------------------------------------------------------------------- 

\subsection{First-order perturbation equations}

We next model the first-order electrostatic (i.e. $B_1 = 0$) perturbations. 
For simplicity, %we shall model the CGL part of the 
the perturbation of the pressure tensor is modelled %as an isotropic, isothermal scalar pressure,
$\tensor p_{e1} = p_{e1} I + \Pi_{e1}$, with
$p_{e1}=n_{e1}T_{e}$
and %$T_{e1}=0$. Here, 
$\Pi_{e1}$ is the perturbed gyroviscous tensor. The zeroth and first-order terms of the divergence of $\Pi_{e}$, relevant in the following, are
given in appendix \ref{app:gyro}. 

In linearized perturbed form, the continuity equation \eqref{0th:cont} for electrons reads
\begin{multline}
    - i \w n_{e1} + n_0 \left( \nabla \cdot \bm u_{e1} + \bm u_{e1} \cdot \nabla \ln n_0 \right)\\
    + \bm u_{e0} \cdot \nabla n_{e1} + n_{e1} \nabla \cdot \bm u_{e0} = \nu_p n_{e1} \ \text{.}
\end{multline}
Defining the  Doppler-shifted frequency  as $\w_e \equiv \w - \bm k \cdot \bm u_{e0}$  
and the normalized number density as $h_{e1} \equiv n_{e1} / n_{e0}$, and rearranging terms, this becomes
\begin{multline}
- i \w_e h_{e1} + \bm u_{e0} \cdot \nabla h_{e1} + i \bm k \cdot \bm u_{e1}
+ \nabla_\perp {u}_{\perp e1} + \nabla_\prl {u}_{\prl e1} \\
- u_{\perp e1} \nabla_\perp \ln B - u_{\prl e1} \nabla_\prl \ln B + \bm u_{e1} \cdot \nabla \ln n_0 =\\
- h_{e1} \left( \frac{\nabla \cdot \left( n_{0} \bm u_{e0} \right)}{n_0} - \nu_p \right) \ \text{,}
\label{1st:cont}
\end{multline}
with the directional derivatives $\nabla_\perp = \bm 1_\perp \cdot \nabla$ and $\nabla_\prl = \bm 1_\prl \cdot \nabla$.
The first-order electron momentum equation \eqref{0th:el:mom}, projected along $\bm 1_\perp,\bm 1_\theta,\bm 1_\prl$, yields, respectively,
\begin{widetext}
    \begin{multline}
        u_{\perp e1} \Big[ - i \w_e + \nu_{e} + \bm u_{e0} \cdot \nabla \ln u_{\perp e1}
        + \nabla_\perp u_{\perp e0}
        - \left( \nabla_\prl \ln B \right) u_{\prl e0} \Big]
        + u_{\theta e1} \w_{ce} 
        + u_{\prl e1} \left[ \nabla_\prl u_{\perp e0} - \left( \nabla_\prl \ln B \right) u_{\perp e0} +
        2 \left( \nabla_\perp \ln B \right) u_{\prl e0} \right] = \\
        - \frac{ i k_\perp p_{e1} + \nabla_\perp p_{e1}}{m_e n_0} - \frac{(\nabla \cdot {\Pi}_{e1})_\perp}{m_e n_0} 
        + \left( \frac{\nabla_\perp p_{e0}}{m_e n_0} + \frac{(\nabla \cdot {\Pi}_{e0})_\perp}{m_e n_0}  \right) h_{e1} + \left( i k_\perp + \nabla_\perp \ln \phi_1 \right) \frac{e \phi_1}{m_e}
        \ \text{,} \label{1st:mom:perp}
    \end{multline}
    \begin{multline}
        u_{\perp e1} \left[ - \w_{ce} + \nabla_\perp u_{\theta e0} \right]
        + u_{\theta e1} \left[ - i \w_e + \nu_{e} + \bm u_{e0} \cdot \nabla \ln {u}_{\theta e1} \right]
        + u_{\prl e1} \nabla_\prl u_{\theta e0} = \\
        - \frac{i k_\theta p_{e1}}{m_e n_0} - \frac{(\nabla \cdot \Pi_{e1})_\theta}{m_e n_0} + \frac{(\nabla \cdot \Pi_{e0})_\theta}{m_e n_0} h_{e1} + i k_\theta \frac{e \phi_1}{m_e}
        \ \text{,} \label{1st:mom:theta}
    \end{multline}
    \begin{multline}
        u_{\perp e1} \left[ - 2 \left( \nabla_\prl \ln B \right) u_{\perp e0} + \nabla_\perp u_{\prl e0} - \left( \nabla_\perp \ln B \right) u_{\prl e0} \right] 
        + u_{\prl e1} \left[ - i \w_e + \nu_{e} + \nabla_\prl u_{\prl e0} - \left( \nabla_\perp \ln B \right) u_{\perp e0} \right] + \bm u_{e0} \cdot \nabla {u}_{\prl e1} = \\
        - \frac{ i k_\prl p_{e1} + \nabla_\prl p_{e1}}{m_e n_0} - \frac{(\nabla \cdot {\Pi}_{e1})_\prl}{m_e n_0} 
        + \left( \frac{\nabla_\prl p_{e0}}{m_e n_0} + \frac{(\nabla \cdot {\Pi}_{e0})_\prl}{m_e n_0}  \right) h_{e1} + \left( i k_\prl + \nabla_\prl \ln \phi_1 \right) \frac{e \phi_1}{m_e}.
        \label{1st:mom:prl}
    \end{multline}
\end{widetext}
Defining $\bm Q_{e} = [ h_{e}, u_{\perp e}, u_{\theta e}, u_{\prl e}]^T$,
Eqs.~\eqref{1st:cont}~to~\eqref{1st:mom:prl} can be cast in matrix form % for the first order form of the amplitudes,
as
%\\\mrip{if HF case is omitted expansion isn't needed}
%
%\begin{equation}
    %{A}_e \bm Q_{e1} = \bm K \frac{e \phi_1}{m_e}
    %\ \text{,} \label{1st:iter}
%\end{equation}
\begin{align}
    \bm Q_{e1} = A_e^{-1} \bm K \frac{e \phi_1}{m_e} \label{1st:iter} \ ,
\end{align}
where $\bm K = \left[ 0, i k_\perp + \nabla_\perp \ln \phi_1, i k_\theta, i k_\prl + \nabla_\prl \ln \phi_1 \right]^T$ and the $4\times4$ matrix ${A}_e$ depends on zeroth-order plasma quantities, their gradients and on the gradients of their first-order perturbations, symbolically written as ${A}_e = {A}_e \left( \w_s, \bm k, \bm Q_{e0}, \nabla \bm Q_{e0}, \nabla \bm Q_{e1} \right)$.
%If inverted, Eq. \eqref{1st:iter} yields the expression
%
%\begin{align}
    %\bm Q_{e1} = A_e^{-1} \bm K \frac{e \phi_1}{m_e} .
%\end{align}
%
%\eag{you can write this one directly as eq. II.10}
 It is important to note here that, for any $Q_{e0}$ and $Q_{e1}$, $O(\rho_e \nabla \ln Q_{e0}) =  O(\rho_e \nabla \ln Q_{e1}) = O(\epsilon)$, thus requiring both of them to be retained in $A_e$ at that order.
Eq. \eqref{1st:iter} is implicit in $\nabla \bm Q_{e1}$:
the recurrence can be resolved by assuming a weakly inhomogeneous plasma, neglecting second order spatial derivatives of any $Q_{e0}$ and $Q_{e1}$:
\begin{multline}
    \frac{\pd \bm Q_{e1}}{\pd x_k} \left( \phi_1, \bm k, \bm Q_{e0}, \nabla \bm Q_{e0}, \nabla \bm Q_{e1} \right) 
    % = 
    % \\
    % \frac{\pd \bm Q_{e1}}{\pd \phi_1} \frac{\pd \phi_1}{\pd x_k}
    % + \sum_{\substack{Q_{e0} \in \{ n_0, \\ T_{e0}, B_0, \bm u_{e0} \}}} \left[ \frac{\pd \bm Q_{e1}}{\pd Q_{e0}} \frac{\pd Q_{e0}}{\pd x_k} + \sum_l \frac{\pd \bm Q_{e1}}{\pd \left( \pd Q_{e0} / \pd x_l \right)} \frac{\pd^2 Q_{e1}}{\pd x_l \pd x_k} \right] + \sum_{\substack{Q_{e1} \in \\ \{ h_{e1}, \bm u_{e1} \}}} \sum_l \frac{\pd \bm Q_{e1}}{\pd \left( \pd Q_{e1} / \pd x_l \right) } \frac{\pd^2 Q_{e1}}{\pd x_l \pd x_k} 
    \simeq\\
    \frac{\pd \bm Q_{e1}}{\pd \phi_1} \frac{\pd \phi_1}{\pd x_k} + 
    \sum_{\substack{Q_{e0} \in \{ n_0, \\ T_{e}, B, \bm u_{e0} \}}}
    \frac{\pd \bm Q_{e1}}{\pd Q_{e0}} \frac{\pd Q_{e0}}{\pd x_k} 
    % = \frac{\pd \bm Q_{e1}}{\pd x_k} \left( \phi_1, \w, \bm k, \bm Q_{e0}, \nabla \bm Q_{e0} \right) 
    \label{1st:iter:dQ1} \ \text{,}
\end{multline}
so that ${A}_e \simeq {A}_e \left( \w, \bm k, \bm Q_{e0}, \nabla \bm Q_{e0}, \nabla \ln \phi_1 \right)$ and $\bm Q_{e1}$ can be expressed as a function of $\phi_1$ and $\nabla \ln \phi_1$ alone.

%----------------------------------------------------------------------------------
%----------------------------------------------------------------------------------

The zeroth-order electron drift velocity $u_{\theta e0}$ is one of the relevant parameters for the expansion. In equilibrium, $u_{\theta e0}$ results from the sum of an $E\times B$ drift and a diamagnetic drift.
%, as shown in [\onlinecite{ramo21}].  
Only the Low-Frequency (LF) limiting cases will be analysed here, where $u_{\theta e0} = O\left( c_{e} \ \epsilon \right)$ and $\w_e = O\left( \w_{ce} \ \epsilon \right)$, relevant for LHDI. The equilibrium electron parallel velocity, on the other hand, will be assumed to be at most of the order of the ion sound speed, $|u_{\prl e0}| \le O \left( c_{s} \right)$, with $c_{s}^2 \equiv {T_{e} /m_i}$ the ion sound speed and $m_i$ the ion mass.

%For the HF case, %presented in appendix \ref{app:HF},
%$A_e^{(0)}$ is invertible, as shown in appendix \ref{app:HF}; however, for the LF case, presented next, $A_e^{(1)}$ needs to be kept to be able to invert equation \eqref{1st:iter}.
 
%----------------------------------------------------------------------------------
%----------------------------------------------------------------------------------

\subsection{Ion solution}

Cold, unmagnetized, singly-charged ions are described by the same Eqs. \eqref{0th:cont} to \eqref{0th:el:mom}, neglecting the magnetic force ($B=0$), pressure tensor ($\tensor p_i=0$), and collisional momentum exchange ($\nu_i=0$). Otherwise, the same procedure as above applies, mutatis mutandi (e.g. substituting $e$ with $i$).
 
Ion motion at equilibrium is of order $u_{i0} \le O\left( c_{s} \right)$, consistently with other studies and observations of MNs \cite{ande69,ahed10f,zhou21a}.
%For ions, $A_i^{(0)}$ is invertible, and the dominant solution $\bm Q_{i1}^{(0)}$ is (neglecting $A_i^{(1)}$ terms)
Keeping only $O(1)$ terms, the ion system yields
 \begin{align}
        \frac{h_{i1}}{{e \phi_1}/{m_i} } = \frac{k^2}{\w_i^2} \ \text{,}  \label{1st:ion:cnt}
\end{align}
\begin{align}
        \frac{\bm u_{i1}}{{e \phi_1}/{m_i} } = \frac{\bm k}{\w_i}  \ \text{.}
\end{align}
%

%----------------------------------------------------------------------------------
%----------------------------------------------------------------------------------

\subsection{Poisson's equation}

%coupling? construction of the dispersion relation?

%Poisson

%quasineutrality

\label{sec:poisson}

Once $h_{i1}$ and $h_{e1}$ have been found as functions of $\phi_1$, a closure relation for $\phi_1$ is needed.
One possibility is, naturally, to employ Poisson's equation. Neglecting second order spatial derivatives:
%, here expanded up to the $O\left( \epsilon \right)$ order:
%
\begin{multline}
    \frac{n_0 e}{\varepsilon_0} \left( h_{i1} - h_{e1} \right) = - \nabla^2 \phi_1 =\\
    \Bigg[ k^2 - \left( 2ik_\perp - \nabla_\perp \ln B \right) \nabla_\perp \ln \phi_1 - \left( \nabla_\perp \ln \phi_1 \right)^2\\
    - \left( 2ik_\prl - \nabla_\prl \ln B \right) \nabla_\prl \ln \phi_1 - \left( \nabla_\prl \ln \phi_1 \right)^2 \Bigg] \phi_1 \ \text{,}
    \label{Poisson}
\end{multline}
with $\varepsilon_0$ the vacuum dielectric permittivity. 
% old formulation can be found after end of document
Alternatively, the system can be closed with the assumption of quasineutrality, which is just the $\varepsilon_0 \to 0$ limit of Eq. \eqref{Poisson}, i.e. $h_{i1} = h_{e1}$. 
This is also the limit found for $\w_{pi}^2 \gg \w_i^2$ in Eq. \eqref{Poisson}, with $\w_{ps}^2 = {{n_0 e^2}/({m_s \varepsilon_0})} \quad \text{($s=i,e$)}$ the plasma  frequency of the $s$-th species, and hence we shall use this for the low frequency dispersion relation.
%\mmm{I may be missing  something but why does this equation retain epsilon0 in the quasineutral limit? it is a parameter that should dissapear}

% can find the discussion on shape of phi in appendix and after document end

%----------------------------------------------------------------------------------
%----------------------------------------------------------------------------------

\section{Low frequency dispersion relation}

\label{sec:disp_rel_deriv}

In the LF case, $\w_e$ and $u_{\theta e0}$ are of order $O\left( \epsilon \right)$ with respect to $\w_{ce}$ and $c_e$ respectively.
Recalling the conditions for the validity our model, we impose the upper limits on the perpendicular wavenumber $k_{\top} \rho_{e} < 1$ and on the parallel wavenumber $|k_\prl c_{e}| < |\w_e|$; this last condition, in the present case, implies $|k_\prl \rho_{e}| \le O\left( \epsilon \right)$.

With the chosen ordering for $u_{\theta e0}$, the inertial convective terms appear as $O\left( \epsilon^2 \right)$ terms and are therefore neglected. In the following we will make use of the following definitions:
\begin{align}
    \w_\perp & \equiv \w_e - \frac{k_\theta c_{e}^2}{2 \w_{ce}} \nabla_\perp \ln \left( \frac{p_{e0}}{B^2} \right) \ \text{,} \\
    \w_\prl & \equiv \w_e - \frac{k_\theta c_{e}^2}{\w_{ce}} \nabla_\perp \ln \left( \frac{p_{e0}}{B^4} \right) \ \text{,} \label{1st:w_prpl}
\end{align}
which are the Doppler-shifted frequencies multiplying the velocity in the directions perpendicular and parallel to the magnetic field respectively, accounting for gyroviscous cancellation\cite{ramo05b}.
Substituting the expression for the gyroviscous tensor from appendix \ref{app:gyro} and keeping only $O(\epsilon)$ terms,
the linearized electron system of Eqs. \eqref{1st:cont} to \eqref{1st:mom:prl} becomes:
\begin{widetext}
    \begin{align}
         - i \w_e h_{e1}  + \left[ i k_\perp + \nabla_\perp \ln \left( \frac{n_0 {u}_{\perp e1}}{B} \right) \right] {u}_{\perp e1} + i k_\theta {u}_{\theta e1}
        + \left[ i k_\prl + \nabla_\prl \ln \left( \frac{n_0 {u}_{\prl e1}}{B} \right) \right] {u}_{\prl e1}= 0 \ \text{.}
    \end{align}
    \begin{multline}
        c_{e}^2 \left[ i k_\perp + \nabla_\perp \ln h_{e1} \right] h_{e1} 
        - i \left[ \w_\perp + i \nu_{e} \right] u_{\perp e1}
        + \left[ \w_{ce} \left( 1 - \frac{k^2 \rho_e^2}{2} \right) + i \frac{k_\perp c_{e}^2}{\w_{ce}} \nabla_\perp \ln \left( \frac{\sqrt{p_{e0}} {u}_{\theta e1}}{B} \right) \right] u_{\theta e1} \\
        +
        i \frac{k_\theta c_{e}^2}{\w_{ce}} \left[ i k_\prl + \nabla_\prl \ln \left( \frac{p_{e0} {u}_{\prl e1}}{B^{5/2}} \right) \right] u_{\prl e1}
        = \left[ i k_\perp + \nabla_\perp \ln \phi_1 \right] \frac{e \phi_1}{m_e} \ \text{,}
    \end{multline}
    \begin{multline}
        i c_{e}^2 k_\theta h_{e1} 
        - \left[ \w_{ce} \left( 1 - \frac{k^2 \rho_e^2}{2} \right) + i \frac{k_\perp c_{e}^2}{\w_{ce}} \nabla_\perp \ln \left( \frac{\sqrt{p_{e0}} {u}_{\perp e1}}{B} \right) \right] u_{\perp e1}
        - i \left[ \w_\perp + i \nu_{e} \right] u_{\theta e1} \\
        - i \frac{k_\perp c_{e}^2}{\w_{ce}} \left[ i k_\prl + \nabla_\prl \ln \left( \frac{p_{e0} {u}_{\prl e1}}{B^{5/2}} \right) \right] u_{\prl e1}
        = i k_\theta \frac{e \phi_1}{m_e} \ \text{,}
    \end{multline}
    \begin{multline}
        c_{e}^2 \left[ i k_\prl + \nabla_\prl \ln h_{e1} \right] h_{e1} 
        - i \frac{k_\theta c_{e}^2}{\w_{ce}} \left[ i k_\prl + \nabla_\prl \ln \left( {\sqrt{B}}{{u}_{\perp e1}} \right) \right] u_{\perp e1}
        +
        i \frac{k_\perp c_{e}^2}{\w_{ce}} \left[ i k_\prl + \nabla_\prl \ln \left( {\sqrt{B}}{{u}_{\theta e1}} \right) \right] u_{\theta e1} \\
        - i \left[ \w_\prl + i \nu_{e} \right] u_{\prl e1}
        = \left[ i k_\prl + \nabla_\prl \ln \phi_1 \right] \frac{e \phi_1}{m_e} \ \text{.}
    \end{multline}
The inverse of the determinant of the matrix $A_e$ of the above linear system,  $D_e$, is
\begin{multline}
    D_e = \w_{ce}^2 \left( \w_\prl + i \nu_{e} \right)
    \Bigg\{ \omega_e \left( 1 - \frac{k^2 \rho_e^2}{2} \right)^2
    - \frac{k_\theta c_{e}^2}{\omega_{ce}} \left( 1 - \frac{k^2 \rho_e^2}{2} \right)  \nabla_\perp \ln \left( \frac{n_0 {u}_{\perp e1}}{B h_{e1}} \right)
    + k^2 \rho_e^2 \left( \omega_\perp + i \nu_{e} \right) - \frac{\W_\prl^2}{\w_\prl + i \nu_{\prl e}}\\
    + \rho_e^2 k_\perp^2 \frac{k_\theta c_{e\perp}^2}{\w_{ce}} \nabla_\perp \ln \left( \frac{{u}_{\perp e1}}{{u}_{\theta e1}} \right) + \w_{e} O(\epsilon)
    %- \frac{R_\prl}{\w_\prl + i \nu_{e\prl}}
    \Bigg\} \ \text{,} \label{1st:De}
\end{multline}
%\newpage
\end{widetext}
where the frequency $\W_\prl$ is defined as 
\begin{multline}
    \W_\parallel^2 = - c_{e}^2 \left[ \left( i k_\prl + \varkappa_\prl + \varkappa_n \right) \left( 1 + \frac{k^2 \rho_e^2}{2} \right) + \varkappa_T k^2 \rho_e^2 \right] \\
    \times \bigg[ \left( i k_\prl + \nabla_\prl \ln h_{e1} \right) \left( 1 - \frac{k^2 \rho_e^2}{2} \right)\\
    + i k_\prl k^2 \rho_e^2
    + \varkappa_\perp k_\theta^2 \rho_e^2 + \varkappa_\theta k_\perp^2 \rho_e^2 \bigg] \ \text{,} \label{1st:R_prl:1}
\end{multline}
with parallel gradients of zeroth-order terms
\begin{align}
\begin{matrix}
    \displaystyle \varkappa_n = \nabla_\prl \ln \left( \frac{n_0}{\sqrt{B}} \right) \ \text{,} &
    \displaystyle \varkappa_T = \nabla_\prl \ln \left( \frac{T_{e}}{\sqrt{B^3}} \right) \ \text{}
\end{matrix}
\end{align}
and parallel gradients of unknown first-order terms
\begin{align}
\begin{matrix}
    \displaystyle \varkappa_\perp = \nabla_\prl \ln \left( {\sqrt{B}}{u_{\perp e1}} \right) \ \text{,} &
    \displaystyle \varkappa_\theta = \nabla_\prl \ln \left( {\sqrt{B}}{u_{\theta e1}} \right) \ \text{,} \\
    \\
    \displaystyle \varkappa_\prl = \nabla_\prl \ln \left( \frac{u_{\prl e1}}{\sqrt{B}} \right) \ \text{.}
    \label{par_grads}
\end{matrix}
\end{align}
%
%\eag{give a number to this equation}
We can now compute $h_{e1}$ keeping only $O(1)$ terms:
\begin{widetext}
\begin{multline}
\frac{h_{e1}}{e \phi_1 / m_e} = \frac{1}{c_e^2} \Biggr\{ 1 - \frac{\w_{ce}^2}{D_e} \left( 1 - \frac{k^2 \rho_e^2}{2} \right) \bigg[ \left( \w_\prl + i \nu_{e} \right)
\times \bigg( \omega_{e} \left( 1 - \frac{k^2 \rho_e^2}{2} \right) + \frac{k_{\theta} c_e^2}{\w_{ce}} \nabla_\perp \ln  \frac{h_{e1}}{\phi_1}  \bigg) \bigg] 
- \w_{ce}^2 \rho_e^2 \bigg[ \left( \varkappa_\prl + \varkappa_n \right) \left( 1 - \frac{k^2 \rho_e^2}{2} \right)\\
+ \left( \varkappa_\prl + \varkappa_n + \varkappa_T \right) k^2 \rho_e^2 \bigg] \nabla_\prl \ln  \frac{h_{e1}}{\phi_1}  \Biggr\} \ \text{.} \label{1st:cont:De}
\end{multline}
\end{widetext}
Now, neglecting second order spatial derivatives, we can simplify the algebra that results by using the weakly inhomogeneous plasma assumption as in Eq.~\eqref{1st:iter:dQ1}, from which  $|\nabla \w_{ce}| \gg |\nabla \w_e|$. Then, from Eq.~\eqref{1st:De} the gradient of $D_e$ can be approximated as
\begin{align}
    \nabla \ln D_e \simeq \nabla \ln \w_{ce}^2 = 2 \nabla \ln B  \ .
\end{align}
The gradient of $\ln h_{e1}$ can be obtained from differentiating Eq. \eqref{1st:cont:De} as
\begin{align}
\nabla \ln h_{e1} \simeq - \nabla \ln c_e^2 + \nabla \ln \phi_1 = \nabla \ln \frac{\phi_1}{T_e} \ \text{.} \label{1st:d_he}
\end{align}
The identity \eqref{1st:d_he}, when substituted into Eq. \eqref{1st:cont:De}, leads to the cancellation of the perpendicular derivatives of $\phi_1$, being $\nabla_\perp \ln \left( {h_{e1}}/{\phi_1} \right) = - \nabla_\perp \ln T_e$.
By defining the following frequencies:
\begin{align}
    \w_{Te} & \equiv - \frac{k_\theta c_{e}^2}{\omega_{ce}} \nabla_\perp \ln T_{e} \ \text{,}\\
    \sigma_\prl^2 &  \equiv c_e^2 \left( 1 - \frac{k^2 \rho_e^2}{2} \right) \bigg[ \left( \varkappa_\prl + \varkappa_n \right) \left( 1 - \frac{k^2 \rho_e^2}{2} \right) \notag\\
    &+ \left( \varkappa_\prl + \varkappa_n + \varkappa_T \right) k^2 \rho_e^2 \bigg] \nabla_\prl \ln T_e \ \text{,}
\end{align}
Eq.~\eqref{1st:cont:De} simplifies to
\begin{multline}
    \frac{h_{e1}}{e \phi_1 / m_e} = 
    \frac{1}{c_e^2} \bigg[ 1 + \frac{\w_{ce}^2}{D_e} \sigma_\prl^2 \\
    - \frac{\w_{ce}^2}{D_e} \left( 1 - \frac{k^2 \rho_e^2}{2} \right)^2 \left( \w_\prl + i \nu_{e} \right) \left( \w_e + \frac{\w_{Te}}{1 - {k^2 \rho_e^2}/{2}}  \right)
     \bigg] \ \text{.} \label{1st:he}
\end{multline}
The velocities yield (keeping only leading $O(1)$ terms)
    \begin{multline}
    \frac{u_{\perp e1}}{e \phi_1 / m_e} = - \frac{i \w_{ce}}{D_e} \left( 1 - \frac{k^2 \rho_e^2}{2} \right) \left( \w_\prl + i \nu_{e} \right) k_{\theta}  \\
    \times \left( \w_e + \frac{\w_{Te}}{1 - {k^2 \rho_e^2}/{2}}  \right) \ \text{} \label{1st:ue:perp}
    \end{multline}
    \begin{multline}
    \frac{ u_{\theta e1}}{e \phi_1 / m_e} = \frac{i \w_{ce}}{D_e} \left( 1 - \frac{k^2 \rho_e^2}{2} \right) \left( \w_\prl + i \nu_{e} \right) k_{\perp}  \\
    \times \left( \w_e + \frac{\w_{Te}}{1 - {k^2 \rho_e^2}/{2}}  \right)
    \ \text{} \label{1st:ue:theta}
    \end{multline}
    \begin{multline}
    \frac{ u_{\prl e1}}{e \phi_1 / m_e} = - \frac{i \w_{ce}^2}{D_e} \left( 1 - \frac{k^2 \rho_e^2}{2} \right)\\
    \times \bigg[  \left( i k_\prl + \nabla_\prl \ln \frac{\phi_1}{T_e} \right) \left( 1 - \frac{k^2 \rho_e^2}{2} \right) + i k_\prl k^2 \rho_e^2 \\
    + \varkappa_\perp k_\theta^2 \rho_e^2
    + \varkappa_\theta k_\perp^2 \rho_e^2 \bigg] \left( \w_e + \frac{\w_{Te}}{1 - {k^2 \rho_e^2}/{2}}  \right)
    \ \text{.} \label{1st:ue:prl}
    \end{multline}
As in Eq.~\eqref{1st:d_he}, Eqs..~\eqref{1st:ue:perp}~to~\eqref{1st:ue:prl} allow us to compute the gradients of each velocity component neglecting second order spatial derivatives:
\begin{align}
    \nabla \ln u_{\perp e1} = \nabla \ln u_{\theta e1} =  \nabla \ln \frac{\phi_1}{B} \ \text{,}
\end{align}
\begin{align}
    \nabla \ln u_{\prl e1} = \nabla \ln \phi_1 \ \text{.}
\end{align}
We can now rewrite $h_{e1}$, ${D_e}$ and $\W_\prl$ as functions of $\phi_1$ and $\nabla \ln \phi_1$.
The above relations imply that the parallel gradients of Eq. \eqref{par_grads} are
%\eag{that the parallel gradients of eq. ... are}
%
\begin{align}
    \varkappa_\perp = \varkappa_\theta = \varkappa_\prl = \nabla_\prl \ln \frac{\phi_1}{\sqrt{B}} \ \text{,}
\end{align}
while $\W_\parallel$ (Eq.~\eqref{1st:R_prl:1}) becomes
\begin{widetext}
\begin{multline}
    \W_\parallel^2 = - c_{e}^2 \bigg\{ \left( i k_\prl + \nabla_\prl \ln  \frac{n_0 \phi_1}{B} - \frac{k^2 \rho_e^2/2}{1+k^2\rho_e^2/2} \nabla_\prl \ln B \right) \left( 1 + \frac{k^2 \rho_e^2}{2} \right) + k^2 \rho_e^2 \nabla_\prl \ln  \frac{T_{e}}{B}  \bigg\} \\
    \times \bigg\{ \left( i k_\prl + \nabla_\prl \ln  \frac{n_0 \phi_1}{B} - \frac{k^2 \rho_e^2/2}{1+k^2\rho_e^2/2} \nabla_\prl \ln B - \nabla_\prl \ln  \frac{p_{e0}}{B} \right)
    \left( 1 + \frac{k^2 \rho_e^2}{2} \right) + k^2 \rho_e^2 \nabla_\prl \ln T_e  \bigg\} \ \text{.} \label{LF:R_prl:2}
\end{multline}
\end{widetext}
The shape of $\nabla \ln \phi_1$ is selected so that the collisonless form of the dispersion relation appears as a real polynomial in $\w_i$. This is done so that the plasma inhomogeneities don't act as spurious sources or sinks of energy:
Appendix \ref{app:grad} offers a more in-depth discussion supporting the above considerations.
%The shape of $\nabla_\prl \ln \phi_1$ is obtained
Then, by setting $\textrm{Im}\{\W_\prl^2\}=0$ we obtain
%, as discussed in section \ref{sec:poisson} (or in Appendix \ref{app:grad}, more in detail):
%
\begin{align}
    \nabla_\prl \ln \frac{n_0 \phi_1}{B} = \frac{1}{2} \nabla_\prl \ln \frac{p_{e0}}{B} - k^2 \rho_e^2 \nabla_\prl \ln T_e \label{LF:phi_shape}
\end{align}
so that $\W_\prl^2$ can be rewritten as
\begin{multline}
    \W_\prl^2 = c_e^2 \left( 1 + \frac{k^2 \rho_e^2}{2} \right)^2\\
    \times \left\{ k_\prl^2 + \frac{1}{4} \left[ \nabla_\prl \ln  \frac{p_{e0}}{B}  - \frac{k^2 \rho_e^2}{1 + k^2 \rho_e^2 / 2} \nabla_\prl \ln B \right]^2 \right\} \label{LF:R_prl:3}  \text{}
\end{multline}
while $\sigma_\prl^2$ becomes
\begin{multline}
    \sigma_\prl^2 = \frac{c_e^2}{2} \left( 1 - \frac{k^4 \rho_e^4}{4} \right) \nabla_\prl \ln T_e \\
    \times
    \left\{  \nabla_\prl \ln  \frac{p_{e0}}{B}  + \frac{k^2 \rho_e^2}{1 + k^2 \rho_e^2 / 2} \left( 3 \nabla_\prl \ln B + k^2 \rho_e^2 \nabla_\prl \ln T_e \right) \right\} \label{LF:r_prl} \ \text{.}
\end{multline}
This procedure is not needed for the shape of $\nabla_\perp \ln \phi_1$, as it was canceled by combining Eqs. \eqref{1st:cont:De} and \eqref{1st:d_he}.
By defining the frequency
\begin{align}
    \w_{Me} \equiv - \frac{k_\theta c_{e}^2}{\omega_{ce}} \nabla_\perp \ln \frac{n_0}{B^2} \ \text{,}
\end{align}
and recalling the definition of $\w_\perp$ from Eq.~\eqref{1st:w_prpl}, the determinant (Eq. \eqref{1st:De}) can finally be rewritten as
\begin{multline}
    D_e = \w_{ce}^2 \left( \w_\prl + i \nu_{e} \right)
    \Bigg\{ \w_{Me} + \w_{Te} + \w_e \left( 1 + \frac{k^4 \rho_e^4}{4} \right)\\
    + i k^2 \rho_e^2 \nu_e
    - \frac{\W_\prl^2}{\w_\prl + i \nu_{e}} \Bigg\}  \text{}
\end{multline}
where we have discarded the $O(\epsilon)$ terms.
Eq.~\eqref{1st:he} becomes
\begin{multline}
    h_{e1} = \frac{e \phi_1}{m_e c_e^2}\\
    \times
    \frac{\displaystyle \w_{Me} + k^2 \rho_e^2 \left( \frac{\w_{Te}}{2} + \w_e + i \nu_e \right) - \frac{\W_\prl^2 - \sigma_\prl^2}{\w_\prl + i \nu_e}}
    {\displaystyle  \w_{Me} + \w_{Te} + \w_e \left( 1 + \frac{k^4 \rho_e^4}{4} \right) + i k^2 \rho_e^2 \nu_e - \frac{\W_\prl^2}{\w_\prl + i \nu_e} } \label{1st:he:1} 
\end{multline}
and by coupling Eq.~\eqref{1st:he:1} with Eq.~\eqref{1st:ion:cnt} through quasineutrality, dividing by $\phi_1$ and keeping only $O ( 1)$ terms, we get the dispersion relation
\begin{multline}
    \frac{k^2 c_s^2}{\w_i^2} = \\
    \frac{\displaystyle \w_{Me} + k^2 \rho_e^2 \left( \frac{\w_{Te}}{2} + \w_e + i \nu_e \right) - \frac{\W_\prl^2 - \sigma_\prl^2}{\w_\prl + i \nu_e}}
    {\displaystyle  \w_{Me} + \w_{Te} + \w_e \left( 1 + \frac{k^4 \rho_e^4}{4} \right) + i k^2 \rho_e^2 \nu_e - \frac{\W_\prl^2}{\w_\prl + i \nu_e} }  \label{LF:disp_rel}
\end{multline}
with $\W_\prl$ containing effects of both parallel wave propagation and gradients. As we will see later, $\sigma_\prl$ is negligible in most cases.

Eq.~\eqref{LF:disp_rel} is clearly akin to Eq.~(31) from [\onlinecite{smol17}], with the additions of magnetic curvature effects, parallel dynamics and plasma inhomogeneities in the $\perp$-$\prl$ meridional plane and serves as the basis for the discussion in the rest of the paper.

%----------------------------------------------------------------------------------

\subsection{Notable limits of the low frequency dispersion relation}

From here, two notable limits can be taken: the dispersion relations for MSHI \cite{smol17} and the MTSI \cite{kral71}, respectively
%\mmm{expand the accronyms unless defined before}, respectively
%
\begin{align}
\frac{k^2 c_s^2}{\omega_{i}^2} &  = \frac{ \displaystyle \w_{Me} }{ \displaystyle \w_{Me} + \w_{Te} + \w_e  }\qquad \text{($k^2 \rho_e^2 \to 0$, $\W_\prl \to 0$) ,} \label{LF:SHI}
\end{align}
and
\begin{align}
\frac{k^2 c_s^2}{\omega_{i}^2}  & = - \frac{c_e^2 k_\prl^2}{ \omega_e^2 - c_e^2 k_\prl^2}\qquad \text{($k^2 \rho_e^2 \to 0$, $\nabla Q_0 \to 0$) ,} \label{LF:MTSI}
\end{align}
both expressed in their quasineutral limit.

On the other hand, if we  assume $\nabla_\prl \ln Q_0=0$, $\nu_e \gg \w_e$ and $k \rho_e \ll 1$, we  recover
\begin{align}
\frac{k^2 c_s^2}{\omega_{i}^2} 
= \frac{ \displaystyle \w_{Me} + i \left( k^2 \rho_e^2 \nu_e + {c_e^2 k_\parallel^2}/{\nu_e} \right) }{ \displaystyle \w_{Me} + \w_{Te} + \w_e + i \left( k^2 \rho_e^2 \nu_e + {c_e^2 k_\parallel^2}/{\nu_e} \right) } \label{LF:univ}
\end{align}
which is the cold, unmagnetized ion limit of Eq.~(13) from [\onlinecite{vran09}].

%\mmm{Anything to conclude from all this? what can we say about the instabilities present in each limit and their mehcanisms? perhaps a statement that all these little cases are just limits of a single instability/dispersion relation? }

The first two limits \eqref{LF:SHI}-\eqref{LF:MTSI} express the two main ingredients for stream and drift-gradient instabilities, which are interspecies drifts due to gradients (in the perpendicular direction) or due to different thermal velocities (in the parallel direction), as will be later shown in section \ref{LF:DG}. The third limit \eqref{LF:univ} shows how the combined presence of drifts and collisions naturally introduces destabilization in the problem by introducing imaginary terms in the dispersion relation, a concept further explored in section \ref{LF:DR}. In summary, these three limits show the building blocks for fluid instabilities that arise in partially magnetized plasmas.

%----------------------------------------------------------------------------------
%----------------------------------------------------------------------------------
%----------------------------------------------------------------------------------

\section{Low frequency instabilities}

\label{sec:LF}

%We now specialize
%the quasineutral limit of
%the dispersion relation obtained in Eq.~\eqref{LF:disp_rel}
%to the MN case, using simulated data from Ref.~[\onlinecite{jime23a}] for the equilibrium plasma quantities and their gradients. 

In general, solutions of the dispersion relation obained in Eq. \eqref{LF:disp_rel}, $\w_i = \w_i \left( \bm k \right)$, come in two regimes: a higher frequency  pair of electron drift waves with $\w_i \sim \bm k \cdot \bm u_{e0}$, and a lower-hybrid pair, with $\w_i \sim k \rho_e \w_{LH}$.
This is a framework similar to that analysed in [\onlinecite{ramo21}] for the \heavy low-frequency modes in the low-drift regime', with an added fourth branch due to the inclusion of parallel dynamics.

In the collisionless limit, the lower-hybrid branches experience destabilization in the form of a reactive instability (pair of complex conjugate solutions). The reason, as anticipated in the introduction, has to do with the existence of `slow' negative energy waves in presence of zeroth-order drifts. In the collisionless case the only energy sinks present in the plasma are `fast' positive energy waves: when the two types of wave have matching frequencies and wavenumbers%\mmm{is this the condition? or the omegas and ks need to match? (stronger condition)}
, an unstable interaction takes place \cite{sagd69}. In the collisional case, no reactive interaction is needed. When inelastic collisions are present in the medium they act as an energy sink, destabilizing the slow wave without the need of coupling with the fast wave.

In general, the presence of collisions widens considerably the parametric instability region, while at the same time lowering  the growth rates of the drift gradient instabilities identified in the collisionless limit. The analysis is therefore focused mostly on drift-gradient (collisionless) instabilities; notwithstanding this, at the end of this section, a paragraph is dedicated to drift-resistive (collisional) instability conditions and how resistive effects alter the collisionless case.

% Two $\w_i$--$k$ plots will be shown in the following paragraphs, using input data from two different points of the MN to show the role of gradients, $k_\prl / k_\theta$ and $k \rho_e$ on the onset of instabilities.

%----------------------------------------------------------------------------------
%----------------------------------------------------------------------------------

\subsection{Drift-gradient instabilities}

\label{LF:DG}

Defining:
\begin{align}
    \Delta \equiv \w_i - \w_e = \bm k \cdot \left( \bm u_{e0} - \bm u_{i0} \right) \ \text{,} \qquad \Delta_\prl \equiv \w_i - \w_\prl \ \text{,}
\end{align}
the former being the Doppler shift between the two species, and the latter being the interspecies Doppler shift accounting for gyroviscous cancellation, Eq.~\eqref{LF:disp_rel} can be hugely simplified by considering either electron drift wave solutions $\w_i^2 \sim \Delta^2 \gg k^2 c_s^2$  or lower-hybrid wave solutions $\w_i^2 \sim k^2 c_s^2 = k^2 \rho_e^2 \w_{LH}^2$. Here we are assuming $|u_{e0}| \gg |c_s|$, as expected in many current-free MNs. Additionally we will assume $\W_\prl^2 \gg \sigma_\prl^2$ being $|\nabla_\prl \ln p_{e0}|\gg|\nabla_\prl \ln T_e|$, as justified in Table \ref{tab:grads}.

For the electron drift wave solutions, Eq.~\eqref{LF:disp_rel} approximately becomes
\begin{multline}
    \w_e^2 + \left( \frac{\w_{Me}}{k^2 \rho_e^2} + \frac{\w_{Te}}{2} + \Delta - \Delta_\prl \right) \w_e \\
    + \left(\Delta - \Delta_\prl\right) \left( \frac{\w_{Me}}{k^2 \rho_e^2} + \frac{\w_{Te}}{2}  \right)
    - \frac{\W_\prl^2}{k^2 \rho_e^2} \simeq 0
\end{multline}
admitting solutions
\begin{multline}
    \w_i = \Delta - \frac{1}{2} \left( \frac{\w_{Me}}{k^2 \rho_e^2} + \frac{\w_{Te}}{2} + \Delta - \Delta_\prl \right) \\
    \times \left[ 1 \pm \frac{\sqrt{\left[ k^2 \rho_e^2 \left(\Delta - \Delta_\prl - \w_{Te}/2 \right) - \w_{Me} \right]^2 + \W_\prl^2 }}{ k^2 \rho_e^2 \left(\Delta - \Delta_\prl + \w_{Te}/2 \right) + \w_{Me} } \right] \ \text{,} \label{LF:hi-wi}
\end{multline}
which are always stable.

Moving to the lower-hybrid branches, following the assumption $k^2 c_s^2 \ll \Delta^2$ and expanding up to the second power of $\w_i / \Delta$, we get
\begin{multline}
    \w_i^2 - \frac{k^2 c_s^2 \w_i}{\Delta_\prl} \frac{\left( \Delta + \Delta_\prl \right) \left( 1 + k^4 \rho_e^4 / 4 \right) -  \w_{Me} - \w_{Te} }{\w_{Me} + k^2 \rho_e^2 \left( \w_{Te}/2 - \Delta \right) + \W_\prl^2 / \Delta_\prl} \\
    - k^2 c_s^2 \frac{\Delta \left( 1 + k^4 \rho_e^4 / 4 \right) - \w_{Me} - \w_{Te} - \W_\prl^2 / \Delta_\prl}{\w_{Me} + k^2 \rho_e^2 \left( \w_{Te}/2 - \Delta \right) + \W_\prl^2 / \Delta_\prl} \simeq 0
\end{multline}
whose solution $\w_i = \w_{ri} + i \gamma$ has real and imaginary parts:
\begin{align}
    \w_{ri} &= k^2 \rho_e^2 \frac{\w_{LH}^2}{2 \Delta_\prl}
    \frac{\left( \Delta + \Delta_\prl \right) \left( 1 + k^4 \rho_e^4 / 4 \right) -  \w_{Me} - \w_{Te} }
    {\w_{Me} + k^2 \rho_e^2 \left( \w_{Te}/2 - \Delta \right) + \W_\prl^2 / \Delta_\prl} \ \text{,} \label{LF:lhdi:wr}
    \\
    \gamma &\simeq \w_{LH} k \rho_e \sqrt{
    \frac{\Delta \left( 1 + k^4 \rho_e^4 / 4 \right) - \w_{Me} - \w_{Te} - \W_\prl^2 / \Delta_\prl}
    {\w_{Me} + k^2 \rho_e^2 \left( \w_{Te}/2 - \Delta \right) + \W_\prl^2 / \Delta_\prl}} \ \text{.} \label{LF:lhdi:gamma}
\end{align}

The expression for $\gamma$ allows us to retrieve the following general instability criterion:
\begin{multline}
    \left[ \Delta \left( 1 + \frac{k^4 \rho_e^4}{4} \right) - \w_{Me} - \w_{Te} - \frac{\W_\prl^2}{\Delta_\prl} \right]\\
    \times
    \left[ \w_{Me} + k^2 \rho_e^2 \left( \frac{\w_{Te}}{2} - \Delta \right) + \frac{\W_\prl^2}{\Delta_\prl} \right] > 0\ \text{.} \label{MN:lhdi:sw}
\end{multline}
This relation underlines how the electron drifts in the ion reference frame are the ones driving the instability, whether they may be due to $a)$ gradients in electrostatic potential, pressure or magnetic field, or $b)$ thermal motion along the magnetic field lines.

%----------------------------------------------------------------------------------

\subsubsection{Long-wavelength limit}

Taking the limit $k \rho_e \to 0$ of relation \eqref{MN:lhdi:sw}, the instability criterion simplifies to:
\begin{align}
    \left[ \Delta - \w_{Me} - \w_{Te} - \frac{\W_\prl^2}{\Delta_\prl} \right] \left[  \w_{Me} + \frac{\W_\prl^2}{\Delta_\prl} \right] > 0 \label{MN:cond:lhi_dw} \ \text{.}
\end{align}
This expression can be easily shown to be a generalization of the MSHI and the MTSI conditions, 
Eqs. \eqref{LF:SHI}-\eqref{LF:MTSI}.
%whose relevant dispersion relations expressed at the end of section~\ref{sec:2}.
Neglecting parallel propagation and gradients, $\W_\prl=0$,
Eq. \eqref{MN:cond:lhi_dw}
yields the MSHI criterion
\begin{align}
    \left( \Delta - \w_{Me} - \w_{Te} \right) \w_{Me} > 0 \ \text{.} \label{LF:SHI:criterion}
\end{align}
Instead, dropping perpendicular inhomogeineities ($\w_{Me}=0$, $\Delta_\prl = \Delta$)
Eq. \eqref{MN:cond:lhi_dw}
yields the MTSI condition
\begin{align}
    \Delta^2>\W_\prl^2 \ \text{.}
\end{align}
%

%----------------------------------------------------------------------------------

\subsubsection{Finite Larmor radius effects}

As $k \rho_e$ becomes non-negligible, rearranging the relation \eqref{MN:lhdi:sw} yields a finite Larmor radius correction to the criterion in \eqref{MN:cond:lhi_dw}:
\begin{multline}
    \left[ \Delta - \w_{Me} - \w_{Te} - \frac{\W_\prl^2}{\Delta_\prl} \right] \left[ \w_{Me} + \frac{\W_\prl^2}{\Delta_\prl} \right] > \\
    - k^2 \rho_e^2 \Bigg[ \left( \frac{\w_{Te}}{2} - \Delta \right) \left( \Delta - \w_{Me} - \frac{\W_\prl^2}{\Delta_\prl} \right) + \\
    \Delta \left( \w_{Me} + \frac{\W_\prl^2}{\Delta_\prl} \right) \frac{k^2 \rho_e^2}{4}
    + \Delta \left( \frac{\w_{Te}}{2} - \Delta \right) \frac{k^4 \rho_e^4}{2} \Bigg] \ \text{,} \label{MN:cond:lhi_lw}
\end{multline}
implying that even if the condition for long-wavelength instabilities is not respected (for instance, left-hand side of \eqref{MN:cond:lhi_lw} $< 0$), we can still have an onset for larger values of the normalized wavenumber $k \rho_e$. On the other hand, if the long-wavelength criterion from \eqref{MN:cond:lhi_dw} is satisfied, higher values of $k^2 \rho_e^2$ might cause $\w_i$ to become real, quenching the instability.

%----------------------------------------------------------------------------------
%----------------------------------------------------------------------------------

\subsection{Drift-resistive instabilities}

\label{LF:DR}

When  collisional effects are included, the dispersion relation becomes a complex polynomial in $\w_i$, yielding in general complex solutions as a result.

Using analytical continuation:
\begin{align}
    0 \simeq \frac{\pd f}{\pd \w_i} \delta \w_i + \frac{\pd f}{\pd \nu_e} \nu_e \ \text{,}
\end{align}
yields the growth rate due to destabilization from dissipative forces:
\begin{align}
    \delta \w_i = - \frac{\pd f / \pd \nu_e}{\pd f / \pd \w_i} \bigg |_{\nu_e = 0} \nu_e \ .\label{DR:cont}
\end{align}
Computing the partial derivatives, and substituting them into Eq.~\eqref{DR:cont}:
\begin{multline}
    \delta \w_i =
    - i \nu_e \displaystyle \left( \w_i^2 - k^2 c_s^2 \right) \left( k^2 \rho_e^2 +
    \frac{\W_\prl^2}{\w_\prl^2} \right) \\
    \times \bigg\{\displaystyle 2 \w_i \left[ \w_{Me} + k^2 \rho_e^2 \left( \frac{\w_{Te}}{2} + \w_e \right) - \frac{\W_\prl^2}{\w_\prl}  \right]\\
    + \left( k^2 \rho_e^2 + \frac{\W_\prl^2}{\w_\prl^2} \right) \left( \w_i^2 - k^2 c_s^2 \right) - k^2 c_s^2 \left( 1 - \frac{k^2 \rho_e^2}{2} \right)^2 \bigg\}^{-1} \label{DR:dw} \ \text{.}
\end{multline}
For our high-drift case of interest, long wavelength ($k \rho_e \to 0$) lower-hybrid drift waves $|\w_i| \sim k c_s \ll \Delta$ can be shown to be destabilized when
\begin{align}
    \w_i \w_{Me} (\w_i^2 - k^2 c_s^2) < 0 ,
\end{align}
which is always true for one of the two branches.
It can also be shown through Eq. \eqref{DR:dw} that drift gradient instability growth rates are in general lowered by the added dissipation. From this, we can state that while collisions offer a more general destabilization criterion, they also tend to reduce unstable growths excited by gradients alone.

%----------------------------------------------------------------------------------
%----------------------------------------------------------------------------------
%----------------------------------------------------------------------------------

\section{Application to a magnetic nozzle plume}

\label{sec:AD}

\begin{table*}
    \centering
    \begin{tabular}{c|c|c|c|c|c|c|c|c|c|c}
        \hline
        Points & $z$ & $r$  & $\w_{LH}$ & $c_s$ &  $u_{\theta e0}$  & $\nabla_\perp \ln n_0$  & $\nabla_\perp \ln B$  & $\nabla_\perp \ln T_e$  & $\nabla_\prl \ln \left( {p_{e0}}/{B} \right)$  & $\nabla_\prl \ln T_e$  \\
        & $ $ [cm] & [cm] & [$10^6$ s$^{-1}$] & [$10^3$ m/s] & [$10^5$ m/s] & [m$^{-1}$] & [m$^{-1}$] & [m$^{-1}$] & [m$^{-1}$] & [m$^{-1}$] \\
        \hline
        A & $4.5$ & $3.4$ & $8.1$ & $1.9$ & $1.7$ & $-39.4$ & $7.5$ & $-2.5$ & $0.4$ & $-0.0$ \\
        B & $8.4$ & $2.5$ & $3.2$ & $1.9$ & $4.5$ & $-98.7$ & $2.5$ & $3.2$ & $-1.9$ & $-0.2$ \\
        C & $22.2$ & $8.3$ & $0.2$ & $1.8$ & $12.8$ & $-32.3$ & $0.6$ & $1.3$ & $-3.4$ & $-0.6$ \\
        \hline
    \end{tabular}
    \caption{%\mmm{call them ABC in a first column; add these text labels near the points of  fig 1 as well; use these labels to refer to the points inside the text}
    Coordinates, zeroth-order plasma quantities and gradients at the three reference points indicated in figure \ref{imag:pedro}.}
    \label{tab:grads}
\end{table*}

\begin{figure}[h]
    \centering
    \includegraphics[width=.4\textwidth]{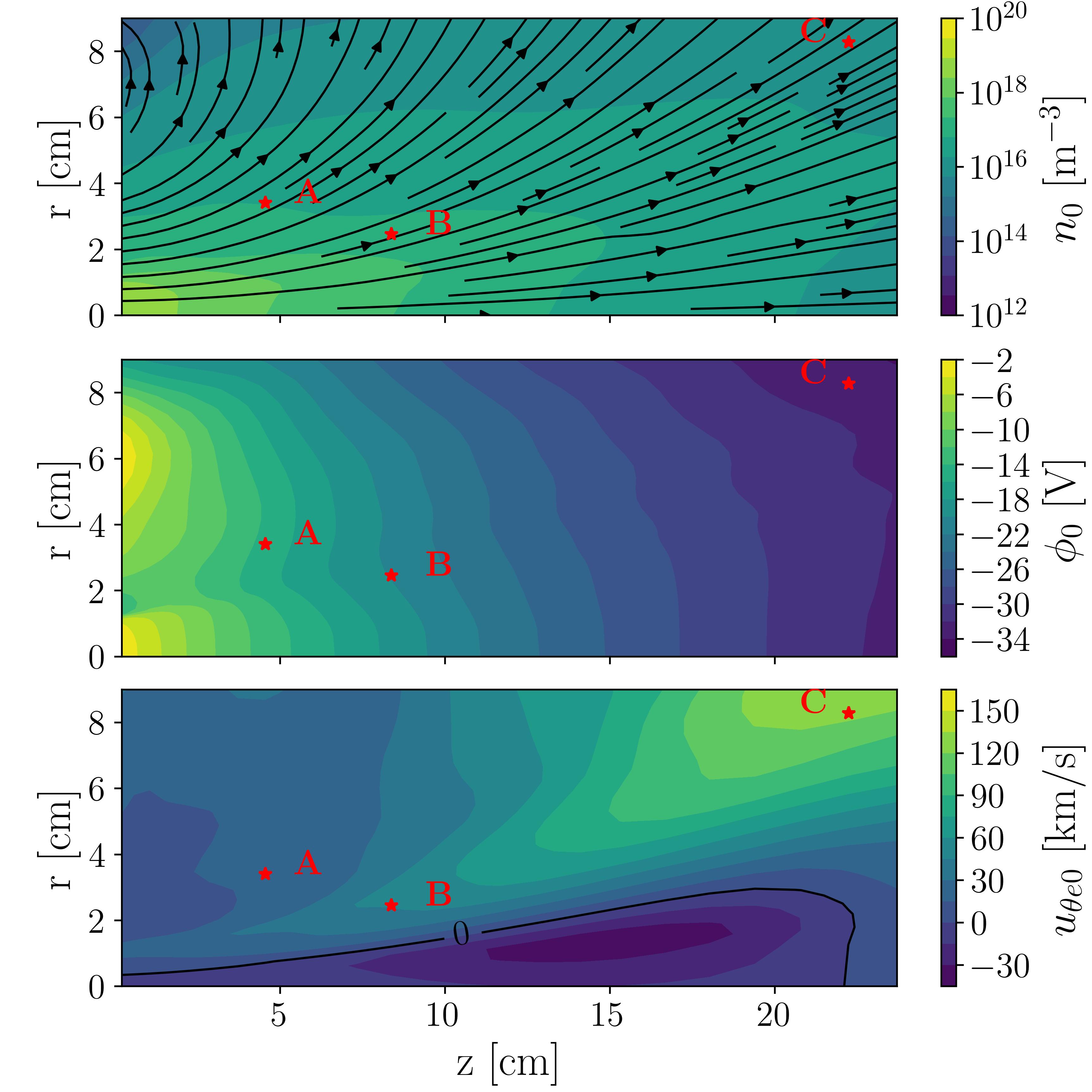}
    \caption{Contour plots for $n_0$, $\phi_0$ and $u_{\theta e0}$, for the simulation data of the plasma expansion in the magnetic nozzle of a helicon plasma thrsuter, from Ref.~[\onlinecite{jime23a}]. %\mmm{superpon B lines en al menos el primer plot}.
    Red dots indicate the three points for which a dedicated $\w_i$-$k$ analysis is presented in subsequent figures. In the plot of $n_0$, $\bm B$ streamlines have been superimposed.}
    \label{imag:pedro}
\end{figure}

As a means of illustration, numerical solutions of \eqref{LF:disp_rel} are obtained for each point in the entire 2D map shown in Figure \ref{imag:pedro}, stemming from the simulations of the MN of a helicon plasma thruster reported in [\onlinecite{jime23a}].
%\mmm{introduce first. Have a paragraph describing the main conditions and assumptions in those simulations, and the cite to the paper.}
The simulation data was obtained through a $2$D hybrid code, HYPHEN-EPT, with electrons modelled as a magnetized diffusive fluid, while the heavy species are simulated through a PIC formulation\cite{zhou22a,pera22a}, and wavefields are solved in the frequency domain using a cold-plasma-wave model \cite{jime23a}. What we show here is a subset of the entire simulation domain, the latter including the plume up to a length of $40$ cm, a radius of $18$ cm, and the ionization chamber as well.

The region shown in Figure \ref{imag:pedro} only covers the MN region directly downstream of the simulated HPT, whose exit is situated at the axial coordinate $z=0$ and extends from $r=0$ up to the radial coordinate $r=1.25$ cm.
Three representative points have been chosen from the shown MN region, two in the near-plume and one in the far-plume part of the discharge, to show three different $\w \left( \bm k \right)$ trends for a drift-gradient instability in the laboratory frame, with $\w = \w_i + \bm k \cdot \bm u_{i0}$.
Point A presents gradients oriented for the most part in the perpendicular direction, with parallel gradients which are of the $O(c_s / c_e)$ order with respect to the perpendicular ones, and $\nabla_\perp \phi \nabla_\perp (n_0 / B^2) <0$, a condition necessary for MSHI as expressed in Eq. \eqref{LF:SHI:criterion};
point B still presents mainly perpendicular gradients but with $\nabla_\perp \phi \nabla_\perp (n_0 / B^2) >0$;
point C, on the other hand, presents parallel gradients which are of order greater than $O(c_s / c_e)$ with respect to the perpendicular ones, thus making the parallel dynamics effect the dominant ones in the dispersion relation.
Each location shows a different evolution of the growth rate $\gamma$ as a function of $k_\theta$, starting from the triggering of a quasi-MSHI in the first case and ending with a short-wavelength instability, mostly driven by parallel dynamics, in the third case. Table \ref{tab:grads} shows their coordinates, and associated zeroth-order plasma quantities and gradients.

%----------------------------------------------------------------------
\begin{figure}[h]
    \centering
    \includegraphics[width=.3\textwidth]{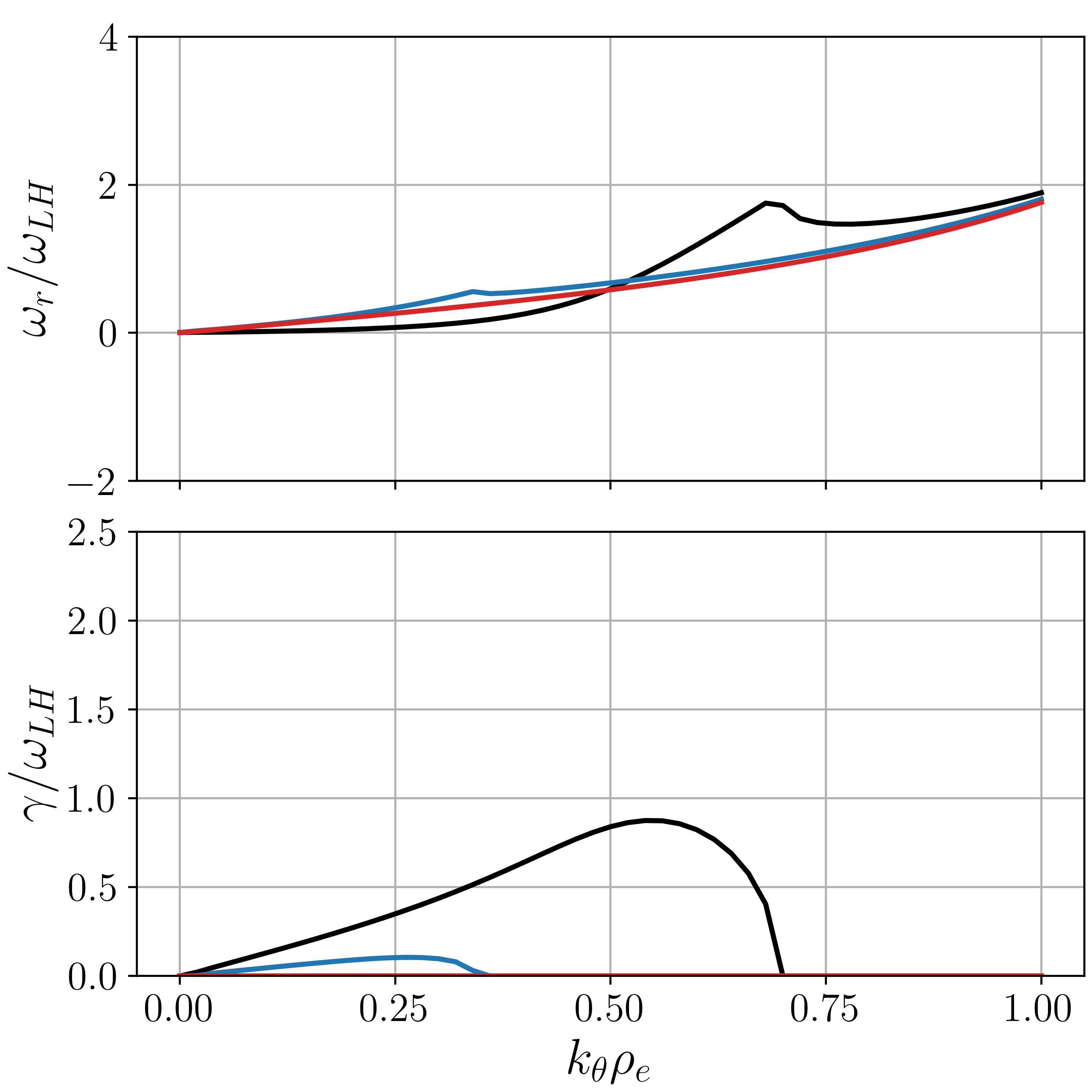}
    \caption{%\mmm{Indicate the point at which you take the plasma properties, and link to the ¿table? with the plasma properties themselves. }
    Real frequency (top figure) and growth rate (bottom figure) for the long-wavelength destabilization of the lower-hybrid branch for the point A of coordinates $(z,r) = (4.5,3.4)$ cm. %\mmm{use label}.
    Black lines are for  ${k_\prl} = 0$; blue lines for $ \displaystyle {k_\prl}/{k_\theta} = 4 \cdot 10^{-3}$; red lines for $ \displaystyle {k_\prl}/{k_\theta} = 8 \cdot 10^{-3}$. 
    %\mmm{one may wonder what happens at other kparallel...}
    Relevant zeroth-order plasma quantities and gradients are shown in Table \ref{tab:grads}.}
    \label{imag:MN:lhdi:lw}
\end{figure}
%----------------------------------------------------------------------

%\mmm{llevar estos plots a sección 'application..'}
%A $\w_i$--$k$ plot is presented in figure~\ref{imag:MN:lhdi:lw}, where real and imaginary parts of $\w_i$ are plotted for $k_\prl/k_\theta = k_\perp=0$ and normalized with $\w_{LH}$. The equilibrium data has been obtained from 
Figure~\ref{imag:MN:lhdi:lw} presents the real and imaginary parts of the solution $\w \left( \bm k \right)$ for the near-plume point A with axial and radial coordinates $(z,r) = (4.5,3.4)$ cm.
For $k_\prl / k_\theta = k_\perp / k_\theta = 0$ it presents an instability mainly driven by perpendicular dynamics, respecting the MSHI criterion from Eq. \eqref{LF:SHI:criterion} and developing mostly in long-wavelength regime, $k \rho_e \ll 1$.
The instability quenches for small values of the ratio $k_\prl / k_\theta$; the effect of $k_\perp$ is, on the other hand, negligible for values of $k_\perp/k_\theta$ of interest, leaving the shape of the solution almost unchanged. For this very reason, in this point and in the following ones, the solutions have been plotted with $k_\perp / k_\theta = 0$.
%\mmm{values of properties and their gradients should be given some where for these chosen points. Table?} 

%----------------------------------------------------------------------
\begin{figure}[h]
    \centering
    \includegraphics[width=.3\textwidth]{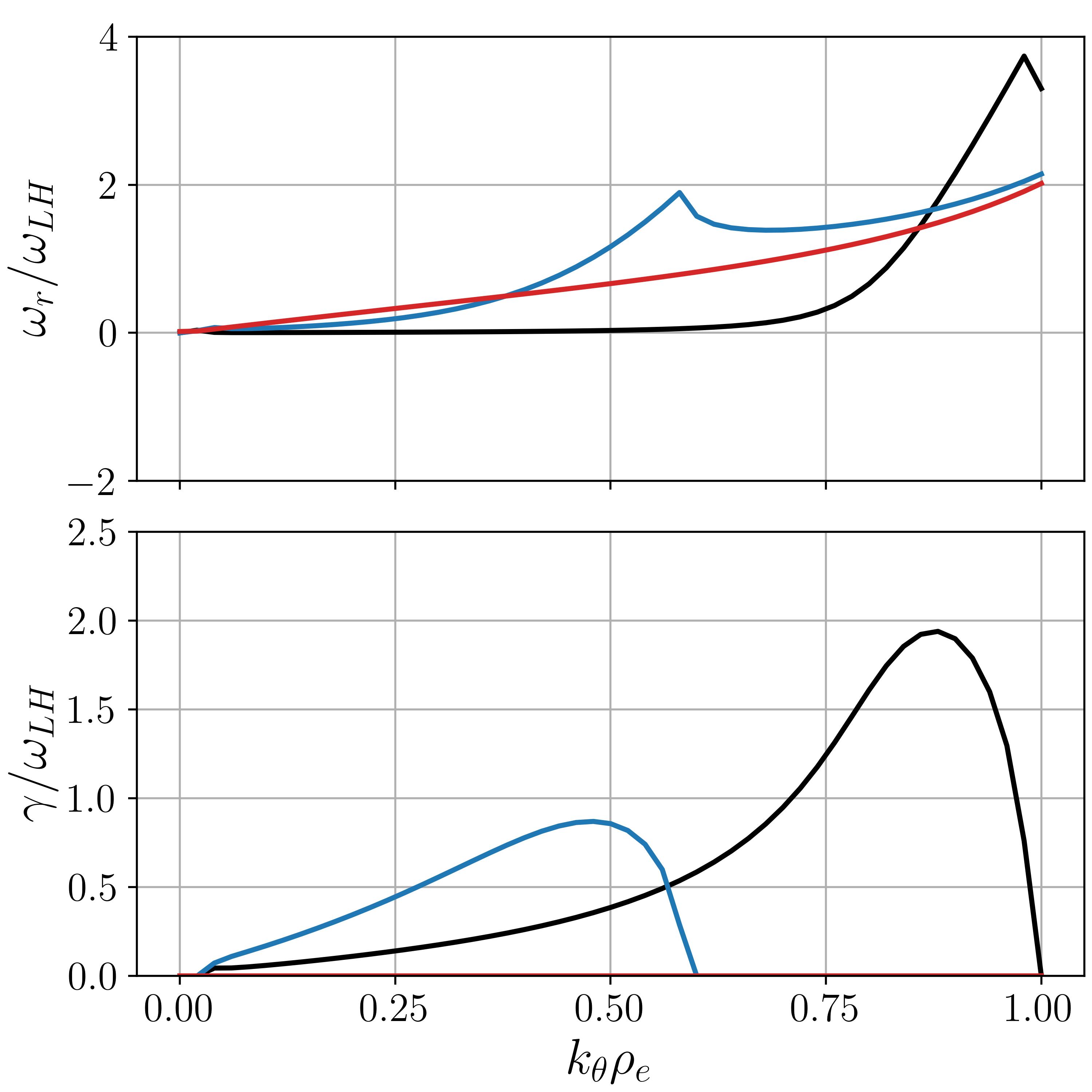}
    \caption{Real frequency (top figure) and growth rate (bottom figure) for the destabilized lower-hybrid branch for the point B of coordinates $(z,r) = (8.4,2.5)$ cm, %\mmm{label}
    with onset in the long-wavelength regime and peak in the short-wavelength regime.
    Black lines are for  ${k_\prl} = 0$; blue lines for $ \displaystyle {k_\prl}/{k_\theta} = 2 \cdot 10^{-2}$; red lines for $ \displaystyle {k_\prl}/{k_\theta} = 4 \cdot 10^{-2}$.
    Relevant zeroth-order plasma quantities and gradients are shown in Table \ref{tab:grads}.}
\label{imag:MN:mid_growth}
\end{figure}
%----------------------------------------------------------------------

Figure~\ref{imag:MN:mid_growth} shows the destabilized lower-hybrid branch for the point B with $(z,r) = (8.4,2.5)$ cm. In this case, the MSHI is no longer respected as the perpendicular components of the electric field and density gradient have different sign, so that the inclusion of parallel dynamics is responsible for the destabilization of the lower-hybrid branch. The instability develops at a slight larger $k \rho_e$ but still in the long-wavelength regime. However its peak is reached in the short-wavelength, at $k \rho_e = O\left( 1 \right)$. Once again, $|k_\prl|>0$ has a stabilizing effect.
%\mmm{justify why you pick these two points. Data in table. It is shocking that this is so different from the previous figure. Is it ok? Would it make sense to show an additional case which is midway between the two?}

%----------------------------------------------------------------------
\begin{figure}[h]
    \centering
    \includegraphics[width=.3\textwidth]{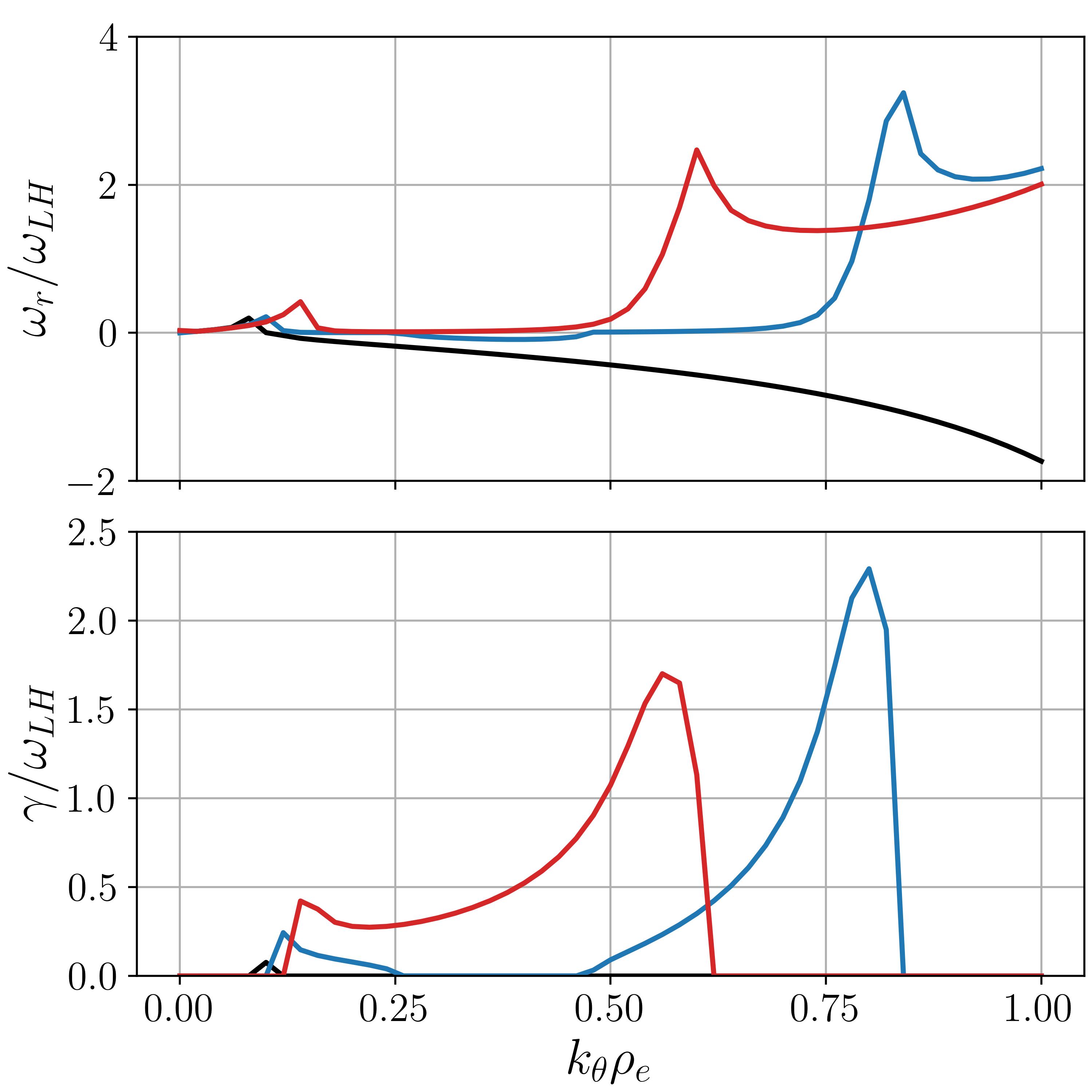}
    \caption{Real frequency (top figure) and growth rate (bottom figure) for the short-wavelength destabilization of the lower-hybrid branch for the point C of coordinates $(z,r) = (22.2,8.3)$ cm .%\mmm{label}.
    Black lines are for  ${k_\prl} = 0$; blue lines for $ \displaystyle {k_\prl}/{k_\theta} = 1.6 \cdot 10^{-1}$; red lines for $ \displaystyle {k_\prl}/{k_\theta} = 2 \cdot 10^{-1}$.
    Solutions with $\w_{r}<0$ are equivalent to solutions with $\w_{r}>0$ but opposite sign of $k_\theta$. 
    Relevant zeroth-order plasma quantities and gradients are shown in Table \ref{tab:grads}.}
\label{imag:MN:FLRI}
\end{figure}
%----------------------------------------------------------------------

Figure~\ref{imag:MN:FLRI} shows the ($\w_i$,$k$) plot of the unstable lower-hybrid branch at the point C with $(z,r) = (22.2,8.3)$ cm.
%\mmm{justify why you pick these two points. Data in table. It is shocking that this is so different from the previous figure. Is it ok? Would it make sense to show an additional case which is midway between the two?}
In this case, far from the MSHI condition and dominated by parallel dynamics, the term $\W_\prl^2 / \Delta_\prl$ for $k_\prl / k_\theta = 0$ (black line) gets cancelled by the large Doppler shift $\Delta \sim k_\theta u_{\theta e0}$, as shown by the small growth rate peak. The instability is then driven by finite parallel propagation $|k_\prl / k_\theta| > 0$, creating a separate short-wavelength onset region as the ratio $|k_\prl / k_\theta|$ grows (blue line). Eventually, the long and short-wavelength onset regions collapse into a single one for larger values of said ratio (red line).

%----------------------------------------------------------------------
\begin{figure*}[t]
    \centering
    \includegraphics[width=.8\textwidth]{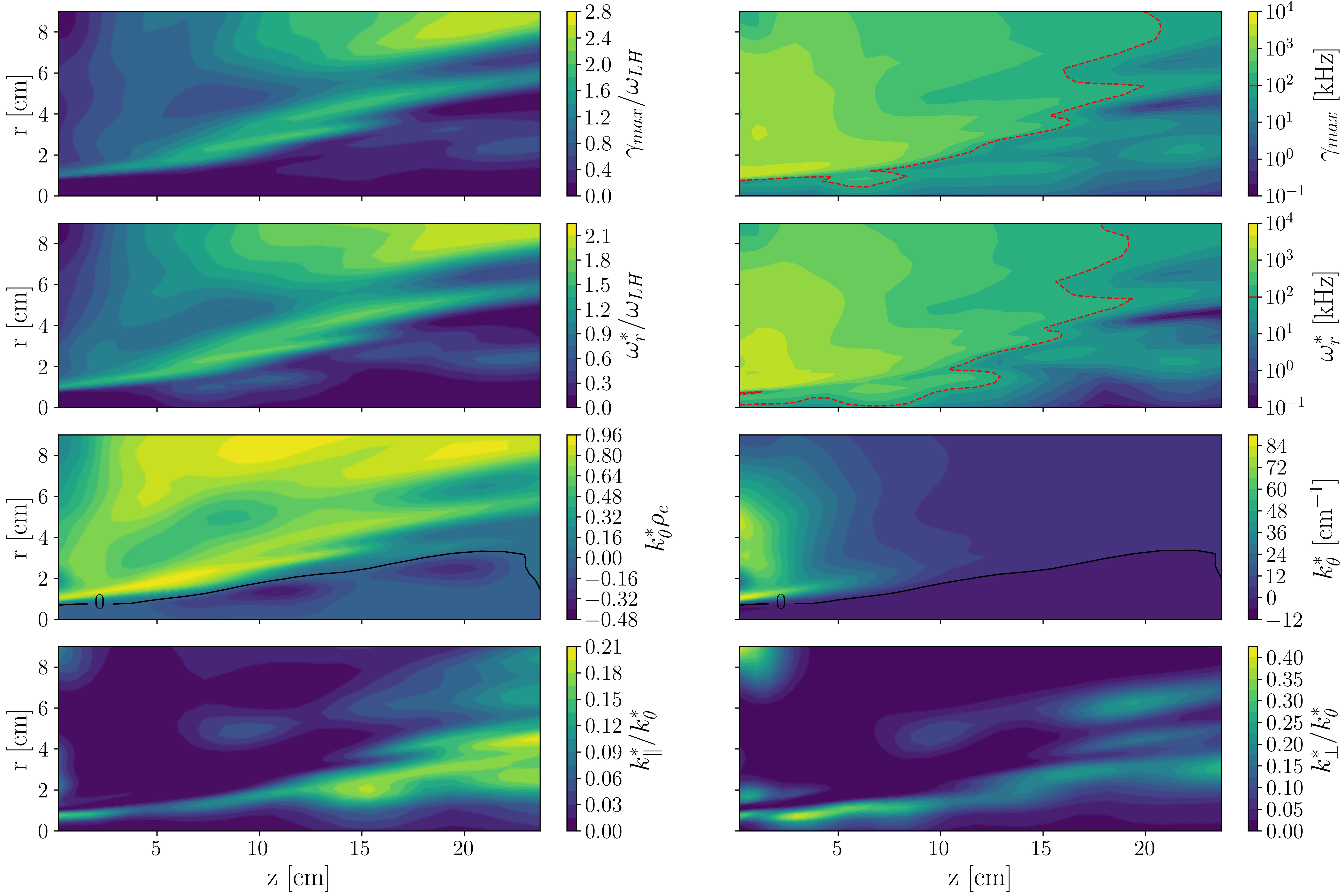}
    \caption{
    %\mmm{asterisks}
    $2$D maps of $\gamma_{max}$, $\w_{r}^*$, $\bm k^*_\theta$ and $k^*_\perp$ resulting from a point-wise local analysis across the MN plume region in the collisionless limit.
    %\mmm{would it be interesting to plot $m_\theta$ instead of $k_\theta?$}
    The dashed red line in the plots of $\gamma_{max}$ and $\w_{r}^*$ represents the $100$ kHz line.
   %\eag{the units should not be in italics (in all figures)} 
    }
    \label{imag:MN:2D_MN_rate}
\end{figure*} 
%----------------------------------------------------------------------
\begin{figure*}[b]
    \centering
    \includegraphics[width=.8\textwidth]{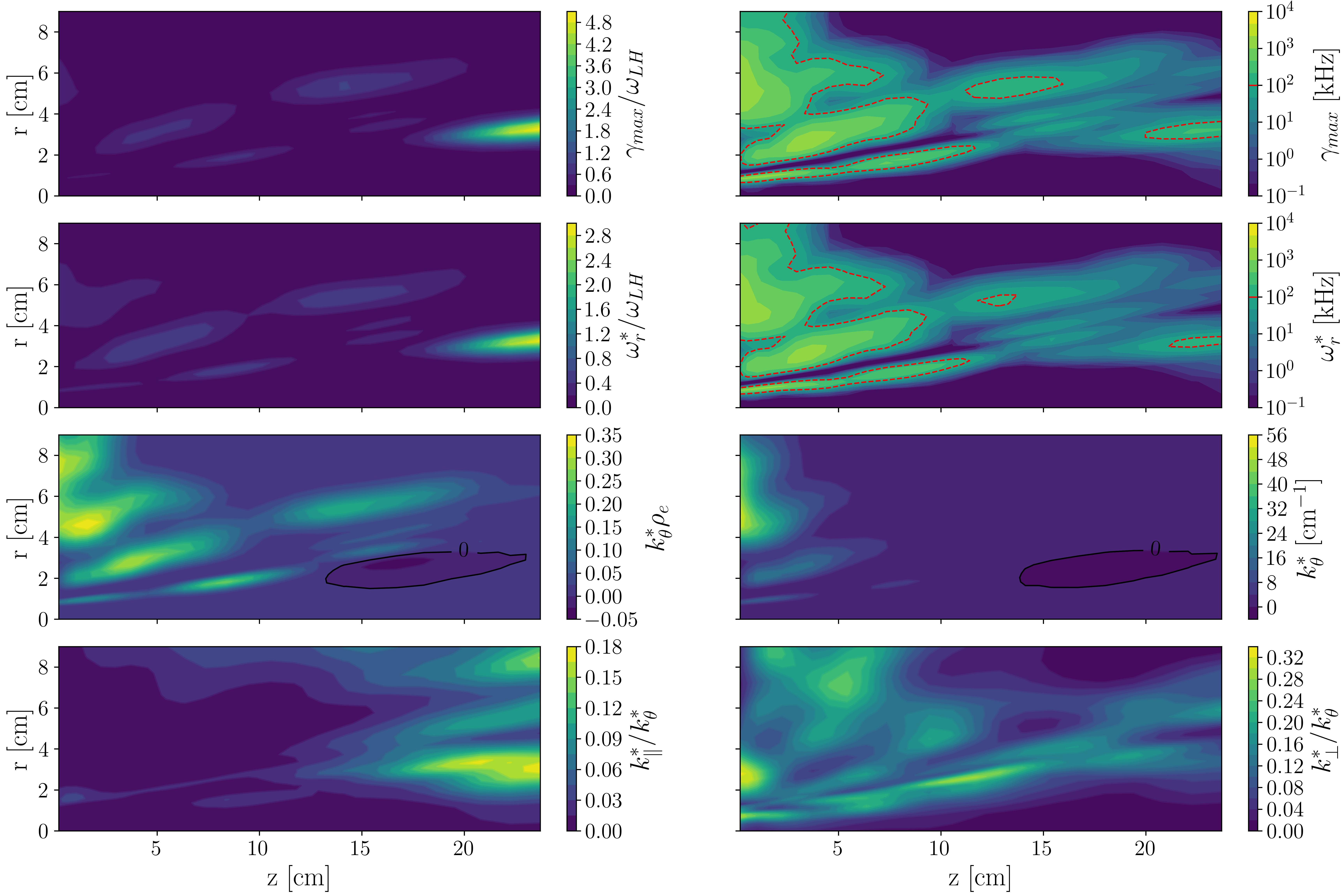}
    \caption{
    %\mmm{I wouldnt change scales wrt previous figure. Mimic caption from previous fig}
    Similar to figure \ref{imag:MN:2D_MN_rate}, but removing parallel gradient effects in the dispersion relation.}
    \label{imag:MN:2D_MN_rate_noprl}
\end{figure*}
%----------------------------------------------------------------------
\begin{figure*}[t]
    \centering
    \includegraphics[width=.8\textwidth]{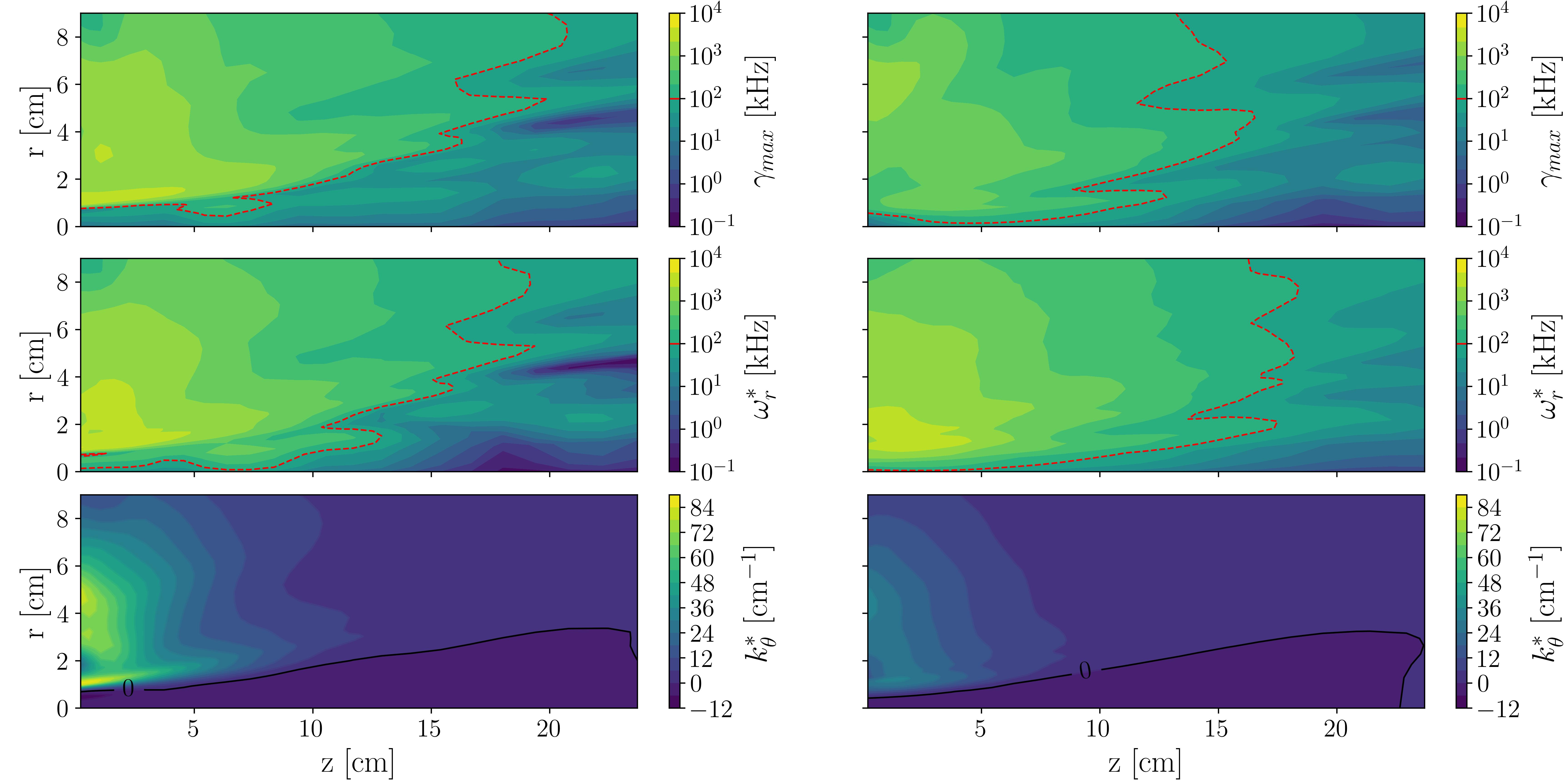}
    \caption{
    %\mmm{would it make more sense to have this after fig 4? }
    Similar to figure \ref{imag:MN:2D_MN_rate}, but including collisions. Left column with $\nu_e$ obtained directly from simulation; right column with $\nu_e $ artificially increased by a factor of $ 10^4$
    %\mmm{either you increase this factor further to show differences, or you leave left column only}.
    .}
    \label{imag:MN:2D_MN_rate_col}
\end{figure*}
%----------------------------------------------------------------------

These three cases have been analyzed to apply the concepts and criteria developed in section \ref{sec:LF} on a relevant configuration. To summarize, various regimes exist in the MN, and fluid waves can be destabilized by either perpendicular gradients, parallel dynamics (gradients and/or wave propagation) or a combination of the the above. The sign of the different perpendicular gradients can be used to assess which kind of mechanism is causing the instability to take place. Gyroviscous effects play a role as well, destabilizing or quenching pre-existing instabilities.

We may now move our attention on the study and identification of the most unstable modes developing at each point of a MN, as predicted by our model.
For each point of the map, the maximum growth rate
%$\gamma_{max} = \displaystyle {\max_{\bm k,\omega_r}} \left( \gamma \left( \bm k, \omega_r \right) \right)$
$\gamma_{max} = \displaystyle {\max} \left( \gamma \left( \bm k, \omega_r \right) \right)$
%$\gamma_{max} = {\max_{\bm k,\omega_r}} \left( \gamma \left( \bm k, \omega_r \right) \right)$
has been obtained, with the associated wavenumber and real frequency in the laboratory frame $\bm k^*$, $\omega_{r}^*$
%\mmm{k asterisk}
respecting the conditions $\rho_e k_{\perp,\theta}^* < 1$ and $|k_{\prl}^* c_e|<|\w_e ^*|$. The mode associated with $\gamma = \gamma_{max}$ is then the most unstable one for that particular point of the MN; $2$D maps of $\gamma_{max}$, $\w_{r}^*$ and $\bm k^*$ are shown in Figure~\ref{imag:MN:2D_MN_rate}. The condition needed for our cartesian expansion to be valid, i.e. $|rk_\theta| \gg 1$, will be relaxed to study points close to the axis.

%\mmm{Better: explain first where the larger gammas are located. Then you may indicate that those regions correspond to larger omega and ktheta for max gamma. Last, kparallel for max gamma seems to be relatively small in those regions, while it is large elsewhere}
%First of all, the peaks of $\gamma/\w_{LH}$, $\w_{ir}/\w_{LH}$ and $k_\theta \rho_e$ are mostly found in the same region of the MN, suggesting the dominant role of short wavelength instabilities. This stresses the importance of finite Larmor effects, especially in the vicinity of $k \rho_e \sim 1$.
The regions with larger $\gamma_{max}/\w_{LH}$ are mostly found in those same points where $|u_{\theta e0}|$ reaches its peak values, as it can be seen comparing $\gamma_{max}$ from Figure \ref{imag:MN:FLRI} with the ones from Figures \ref{imag:MN:lhdi:lw} and \ref{imag:MN:mid_growth}. This can be explained by looking at the relation between $\gamma$ and $\Delta$ from Eq. \eqref{LF:lhdi:gamma}, with 
\[\gamma \simeq k c_s \sqrt{\frac{\Delta - \w_{Me} - \W_\prl^2/\Delta_\prl}{\w_{Me} + \W_\prl^2/\Delta_\prl}}. \]
A similar consideration can be done with the plot of $\w_{r}^*$, recalling the expression of $\w_{r}$ from Eq. \eqref{LF:lhdi:wr}, and with the plot of $k_\theta^* \rho_e$, as $\Delta \propto k_\theta$. All these first three plots show very similar trends, as they all share a direct correlation with the Doppler shift $\Delta$ and the zeroth-order electron drift velocity $u_{\theta e0}$.
The ratio $k_\prl^*/k_\theta^*$ is quite negligible in the near-plume region. As we've stated in the discussion of the three ($\w_i$,$k$) plots, in this region the effect of a finite parallel propagation is, in general, to stabilize the wave. This is not true in the 
%$u_{\theta e0}<0$ region and in the far-plume region, as shown in Figure~\ref{imag:MN:FLRI}.
regions where the perpendicular gradients get close to $0$, that is, closer to the axis and in the far-plume region, as as shown in Figure~\ref{imag:MN:FLRI} and in Table \ref{tab:grads}.
The role of $k_\perp^*$, on the other hand, is quite marginal, as we've stated beforehand. It follows a trend more or less similar to that of $k_\prl^*$, peaking in a region of small $\gamma_{max}$.
In those points where $\gamma_{max}$ is reached for $k^* \rho_e = 1$, a kinetic formulation of the problem would be more suitable.

Most of the instabilities and their associated real frequencies fall in the $1$ kHz--$1$ MHz range, decreasing as we move from the near-plume to far-plume region of the discharge. The modes with larger $\gamma_{max}$ have a mainly-azimuthal associated wavenumber, $\bm k^* \simeq k_\theta^* \bm 1_\theta$.

Noticeably, the inclusion of parallel gradients in the dispersion relation plays a major role in the stability of the solution. Their presence allows the onset of instabilities in those points of the MN where the MSHI criterion from Eq. \eqref{LF:SHI:criterion} does not hold,
%\mmm{complex statement, not clear}.
even without $k_\prl \ne 0$.
This statement is further illustrated by the plots of Figure~\ref{imag:MN:2D_MN_rate_noprl}, which have been computed by using a form of Eq. $\eqref{LF:disp_rel}$ \textit{without} parallel gradients, $\nabla_\prl \ln Q = 0$. These plots present a substantially different map for $\gamma_{max}$, with the majority of the instabilities taking place in those regions satisfying the MSHI criterion.
%\mmm{this seems quite important to be dismissed as something minor? also, figure with differing scales is not useful}
The map of $k_{\theta}^*$ shows milder peaks, loosely following the ones of $\gamma_{max}$.
$k_\prl^*$ has a much more prominent role in the onset of instabilities: this is due to the fact that, in the absence of parallel gradients, $\W_\prl^2 \propto k_\prl^2$. In the regions where the MSHI criterion is not satisfied, then, instabilities can only be driven by finite parallel propagation i.e. $k_\prl \ne 0$.
It is therefore clear that parallel gradients of plasma quantities have to be retained in the formulation to consistently include the complete set of mechanisms which lead to linear fluid instabilities of the plasma in the LF regime. The present work, to our knowledge, is the first to include said gradients.

%Parallel dynamics are quite important in the onset of instabilities, allowing non-zero growth rates throughout the entire MN, even in those regions where conditions for SHI are not met. This is highlighted in the plots of Figure~\ref{imag:MN:2D_MN_rate_noprl}, where in the absence of parallel gradients instability peaks coincide with points of $k_\prl \ne 0$.

%\mmm{Put this a bit later? before you discuss parallel gradient effect?}
The effect of collisions on the instability peaks are negligible, as electron-neutral collision frequencies remain in the order $\nu_e \le O\left( \w_{LH} \right)$, barely affecting the $O\left( \w_{LH} \right)$ growth rates. Overall, the $2$D maps of $\gamma_{max}$, $\w_{r}^*$ and $\bm k^*$ remain almost unchanged between the collisionless to the collisional case, as the most unstable modes are of the drift-gradient type. Figure~\ref{imag:MN:2D_MN_rate_col} compares two collisional cases, the first where the collisional frequency is directly obtained from the simulation data, and in the second case the same frequency is multiplied by a factor of $10^4$. Collisions show to have an effect only in those regions where drift gradient instabilities are minor or completely absent, slightly extending the regions of instability and slightly reducing the peak of gradient-driven growth rates.

We conclude this section with a qualitative comparison with the limited available experimental data. Recalling Figure \ref{imag:MN:2D_MN_rate}, we have shown that the most unstable modes predicted by our model consist of mainly-azimuthal waves with associated real oscillation frequency $\w_{r}^*$ in the $1$ kHz--$1$ MHz range.
While Hepner et al.\cite{hepn20b} and Vinci\cite{vinci22e} do find fluctuations in a similar frequency range in the MN of their respective devices, they both describe waves with combined axial-azimuthal propagation (where the presence of a large $k_\prl$ introduces kinetic effects which elude our fluid model).
Indeed, Hepner et al. attribute their observed fluctuations to an LHDI and invoke a kinetic formulation to justify it, involving values of parallel wavenumber comparable to the one in the azimuthal direction.
In the works of both Takahashi et al.\cite{taka22a} and Maddaloni et al.\cite{madd24b}, fluctuations are detected in their respective MNs, which fall in the $10$--$100$ kHz range and consist of mostly azimuthal waves.
Takahashi et al. identifiy their oscillations with azimuthal magnetosonic waves, while Maddaloni et al. present the case that the fluctuation spectrum they observe is due to a nonlinear parametric decay instability of the pump wave at 13.56 MHz, as the conditions for LHDI are not met where said fluctuations are observed. Our work suggests that an alternative explanation to either of them could be that the observed fluctuations are the signature of low frequency, drift-gradient instabilities of the type described above, which can be triggered even in those regions where the MSHI criterion is not satisfied, and the parallel wavenumber is too close to zero to justify the onset of MTSI.

%Among the cited experimental works, our predicted unstable modes seem to more qualitatively describe fluctuations observed by Takahashi et al.\cite{taka22a} and Maddaloni et al.\cite{madd24b}, which fall in the $10$--$100$ kHz range and consist of mostly azimuthal waves.
%
%Hepner justifies the observed fluctuations through  the kinetic formulation of the LHDI, involving values of parallel wavenumber comparable to the one in the azimuthal direction.
%
%Takahashi, on the other hand, identifies the oscillations as azimuthal magnetosonic waves. These, however, are MHD oscillations that are stable when traveling in the direction perpendicular to the magnetic field.
%
%In his work, Maddaloni explains the presence of fluctuations through parametric decay instability, as the conditions for LHDI are not met where said fluctuations are observed. An alternative to this explanation would be considering the parallel-gradient-driven instabilities presented in this work, as azimuthal oscillations are present even in those regions where the MSHI criterion is not satisfied, and the parallel wavenumber is too close to zero to justify the onset of MTSI.

%----------------------------------------------------------------------------------
%----------------------------------------------------------------------------------
%----------------------------------------------------------------------------------

\section{Quasi-linear cross-field transport}

\label{sec:ql}

In the collisionless case, the electron cross-field current at the zeroth order is null. This allows instability-driven electron transport to be a dominant term in this direction. 
%Our study has focused on the linear perturbations of equilibrium plasma quantities in  a MN.
%We now tackle the effect of said linear perturbations on cross-field electron transport. To do so,
To tackle the effect of linear perturbations on cross-field electron transport we quantify, through a limited quasi-linear analysis, when and how the non-linear interaction between these oscillations can produce a non-zero second order term in the electron current 
field.

These second order terms are generally comprised of a quasi-DC axysimmetric part, varying in time as $2 \gamma t$, and a double frequency part.
Since we want to gauge how instabilities affect the equilibrium over time, it is only the quasi-DC part that we are interested in.
For this purpose, we consider our plasma quantities to be comprised of an additional, quasi-DC and axisymmetric second order term, their expression now being 
\begin{multline}
Q\left( \bm x, t \right) = Q_0 \left( \bm x \right) + \frac{1}{2} [Q_1 \left( \bm x \right) \exp\left( i \bm k \cdot \bm x - i \w t \right) + CC]\\
+ Q_2 \left( \bm x\right) \exp(2 \gamma t) + \dots
\end{multline}
We note that this expansion is not uniformly valid for all times; the second-order term will eventually result in a modified equilibrium solution, requiring the redefinition of $Q_0$. In the following, we implicitly restrict our analysis to the time interval of validity of this expansion.

For the cross-field transport, we are interested in the second order, quasi-DC, perpendicular electron flux, %\eag{which}
which is  given by:
\begin{align}
    \left( n_e u_{\perp e} \right)_2 = 
    n_0 \Braket{ u_{\perp e1} h_{e1}^*} \exp(-2 \gamma t) + n_0 u_{\perp e2} \ \text{.} \label{ql:flux}
\end{align}
%\eag{You can take $n_0$ as common factor and elminiate it from the expressions below. You must explain the interpretation to be made of $\exp(2 \gamma t)$}
%
%where we have defined as $\Braket{a_1 b_1^*}$  the quasi-DC part of the real part of the product two generic first order quantities,
%\eag{(alt) Here,  the quasi-DC, real part of the product two generic first order quantities, say $a_1$ and $b_1$, is}
Here,  the quasi-DC, real part of the product two generic first-order quantities, say $a_1$ and $b_1$, is
\begin{align}
    \Braket{a_1 b_1^*} = \frac{ a_1  b_1^* +  a_1^*  b_1}{4}  \exp(2 \gamma t)
    \text{.} \label{ql:AB_1}
\end{align}
%
%having assumed $\gamma < \w_r$.
The $\Braket{u_{\perp e1} h_{e1}^*}$ term on the right-hand side of Eq. \eqref{ql:flux} can be computed by substituting for $u_{\perp e1}$ the expression from \eqref{1st:ue:perp}
\begin{multline}
    u_{\perp e1} = \\
    -
    \frac{\displaystyle \frac{i k_\theta}{\w_{ce}} \left( \w_e + \frac{\w_{Te}}{1-k^2 \rho_e^2/2} \right) \left( 1 - \frac{k^2 \rho_e^2}{2} \right)}
    {\displaystyle\w_e \left( 1 - \frac{k^2 \rho_e^2}{2} \right)^2 + \w_{Me} + k^2 \rho_e^2 \left( \frac{\w_{Te}}{2} + \w_e \right) - \frac{\W_\prl^2}{\w_\prl}} \frac{e \phi_1}{m_e}\ ; \notag
\end{multline}
making use of the LF dispersion relation (Eq. \eqref{LF:disp_rel}), the expression for $ \Braket{u_{\perp e1} h_{e1}^*}$ renders as
\begin{multline}
    \Braket{ u_{\perp e1} h_{e1}^*} = 
    -  \frac{k_\theta / \w_{ce}}{ 1 - k^2 \rho_e^2 / 2 } \Braket{i \left( \frac{e \phi_1}{m_e} - \frac{k^2 c_s^2}{\w_i^2} \frac{e \phi_1}{m_e} \right)  h_{e1}^*} = \\
    c_e \frac{k_\theta \rho_e}{ 1 - k^2 \rho_e^2 / 2 } \left( \frac{k c_s}{\w_{ri}^2 + \gamma^2} \right)^2 \w_{ri} \gamma
    \frac{e^2 |\phi_1|^2}{T_e^2} \exp(2 \gamma t) \ \text{.} \label{ql:F_ue}
\end{multline}
%\eag{The writing of the above sentence can be improved,... and omit $n_0$}
%
%\mmm{Caution. When you take the angle brackets, you take the complex conjugate, affecting iomegat and imtheta. This result is linear with ktheta. Is this correct? should it depend on abs(ktheta)?}
In the last passage, we have used $h_{e1} = h_{i1} = k^2 c_s^2 e \phi_1 / T_e$, from Eq. \eqref{1st:ion:cnt}.
%\eag{wherefrom? should it be $h_{e1}$?}, $\w_i = \w_{ri} + i \gamma$ 
%\eag{obvious} and ${\Braket{ia_1 a_1^*}} = 0$ \eag{obvious}.

\begin{figure}
    \centering
    \includegraphics[width=.4\textwidth]{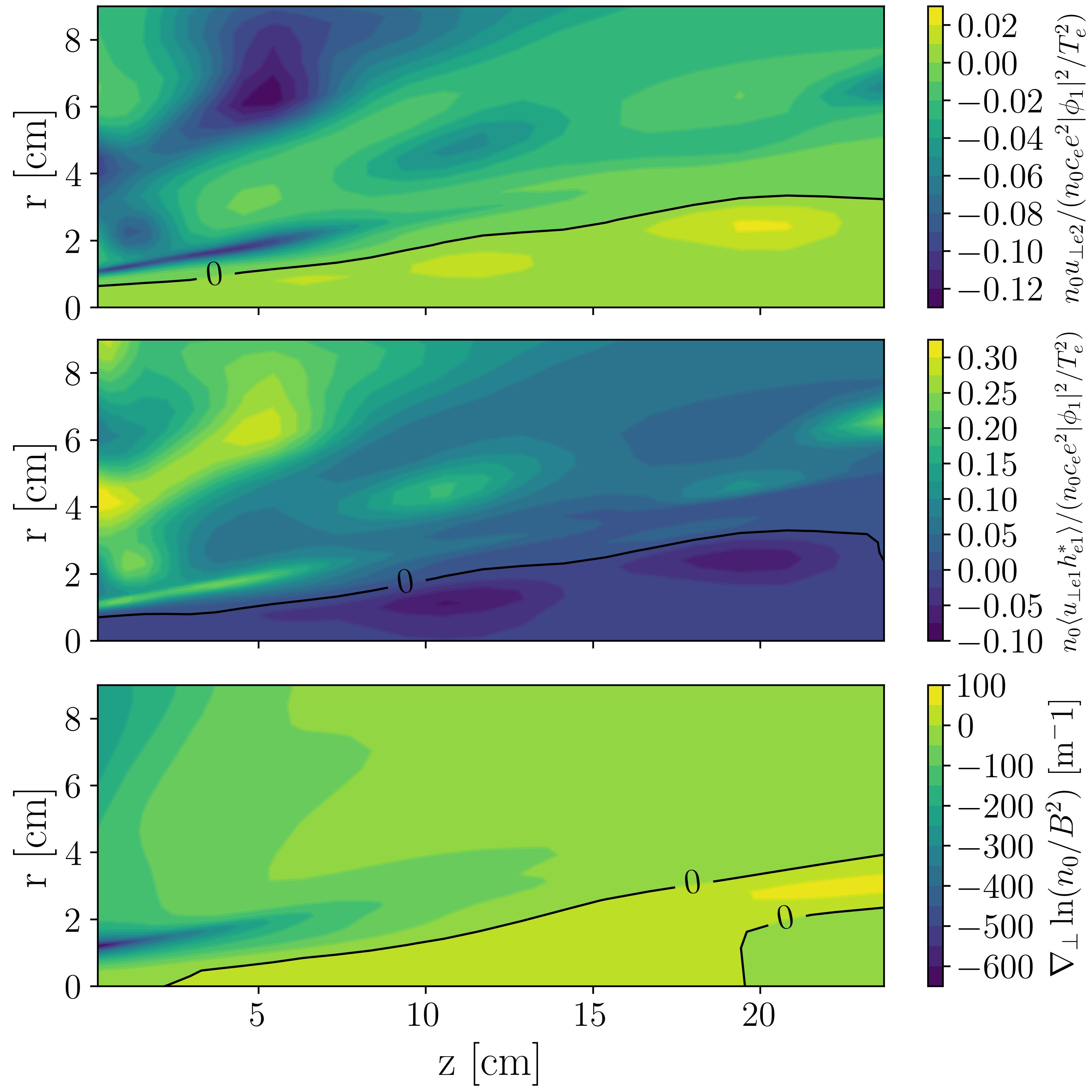}
    \caption{Top: $2$D map of $n_0 u_{\perp e2}$ normalized with respect to the local electron thermal flux $n_0 c_e$ and the non-dimensional wave amplitude $e^2 |\phi_{1}|^2 / T_e ^2$. Middle: $2$D map of $n_0 \Braket{ u_{e\perp1} h_{e1}^*}$ normalized with respect to the local electron thermal flux $n_0 c_e$ and the non-dimensional wave amplitude $e^2 |\phi_{1}|^2 / T_e ^2$. Bottom: $2$D map of the gradient $\nabla_\perp \ln (n_0/B^2)$.}
    \label{fig:ql:u2}
\end{figure}

%\eag{change paragraph} 
The quasi-DC second order cross-field velocity $u_{\perp e2}$ can be obtained from
the second order
%form \eag{expansion?} of \eag{the momentum }
%the 
momentum equation \eqref{0th:el:mom}:
\begin{multline}
    2 \gamma \bm u_{e2}
    + \bm u_{e0} \cdot \nabla \bm u_{e2} + \bm u_{e2} \cdot \nabla \bm u_{e0}  + \Braket{ \bm u_{e1} \cdot  \nabla \bm u_{e1}^*} \exp(-2 \gamma t) = \\
    - \frac{\nabla p_{e2}}{m_e n_0} - \frac{\nabla \cdot \Pi_{e2}}{m_e n_0} + \frac{e \nabla \phi_2}{m_e} + \w_{ce} \bm 1_\prl \times \bm u_{e2} \\
    - \left( \frac{\nabla p_{e0}}{m_e n_0} + \frac{\nabla \cdot \Pi_{e0}}{m_e n_0} \right) \left( \Braket{ h_{e1} h_{e1}^* } \exp(-2 \gamma t) - h_{e2} \right) \\
    + \Braket{ \left(  \frac{\nabla p_{e1} }{m_e n_0} + \frac{ \nabla \cdot \Pi_{e1} }{m_e n_0} \right) h_{e1}^* } \exp(-2 \gamma t)
    \text{,} 
    \label{2nd:el:mom}
\end{multline}
%
%\mmm{careful, this equation should already be for the DC part. So the u1*u1 terms should be in angle brackets I guess. And no CC things then.}
where we have used the expansion $\left(n^{-1} Q\right)_2 = n_0^{-1} \left( Q_2 - Q_1 n_1 + Q_0 n_1^2 n_0^{-1} - Q_0 n_2 \right) \exp(2 \gamma t)$. 
In the LF regime $\gamma = O(\epsilon\ \w_{ce})$, so that the left-hand side of the above equation, %\eag{i.e. the inertia}
i.e. the inertia, is of order $\epsilon$ with respect to the right-hand side. 
%We need to consider the projection of $\w_{ce} \bm 1_\prl \times \bm u_{e2}$ along the azimuthal direction to obtain an expression for $u_{\perp e 2}$.
%\eag{Rewrite: you mean that $u_{\perp e 2}$ is obtained from the magnetic force term in the azimuthal momentum equation  (as we do in 0th order)}
The perpendicular velocity $u_{\perp e 2}$ is then obtained from the magnetic force term in the azimuthal momentum equation.
Using
$\displaystyle \nabla_\theta Q_0 = 0$ and $\nabla_\theta Q_2 = 0$
and Eq. \eqref{gyro:2nd} (yielding $\nabla \cdot \Pi_{e2} = O(n_0 m_e \w_{ce} u_{\perp e 2}\ \epsilon)$), Eq.~\eqref{2nd:el:mom} gets rewritten as
%\mmm{omegace can be very large, so I am unsure if you can really drop it. Also, you dont drop it below:}
%
\begin{multline}
    \w_{ce} u_{\perp 2} \exp(2 \gamma t) =\\
    - \Braket{ \left( i k_\theta c_e^2 h_{e1} + \w_{ce} \frac{k^2 \rho_e^2}{2} u_{e\perp1} \right) h_{e1}^*}  (1 + O(\epsilon))
    \label{2nd:el:mom:theta_1} \ \text{.}
\end{multline}
%
%\eag{Is there any remain here from gyroviscosity?} \mrip{written below}
Having assumed isothermal perturbations, $T_{e1} = 0$, a reasonable assumption for low frequency oscillations \cite{bell21a}, the contribution of pressure to second order cross-field flow is null. This only leaves the contribution from the product between the first-order divergence of the gyroviscous tensor from Eq. \eqref{gyro:theta} and the first-order number density. This term is the second one appearing in the right hand side of Eq. \eqref{2nd:el:mom:theta_1}, and unlike inertial terms yields  a contribution of order $O(k^2 \rho_e^2)$ with respect to the magnetic force.
By making use of Eq. \eqref{ql:F_ue} and $c_s = \rho_e \w_{LH}$ and neglecting $O(\epsilon)$ terms, we obtain the second order electron cross-field velocity
\begin{multline}
    u_{\perp e2} =
    - c_e \rho_e \w_{LH}^2 \frac{k^4 \rho_e^4 / 2}{1 - k^2 \rho_e^2 / 2}
    \frac{k_\theta \w_{ri} \gamma}{\left(\w_{ri}^2 + \gamma^2\right)^2} 
    \frac{e^2 |\phi_1|^2}{T_e^2} \ \text{.} \label{2nd:el:uperp2}
\end{multline}
%
%\eag{what about $h_{e1}$? Why don't you add $\w_{ri}^2+ \gamma^2$ above and you simplify?  I would use $|\phi_1|^2$}
%\mmm{The derivation above is a bit convoluted/disorganized. Firstly, it should be clear whether our goal is to model the full flux in eq 68, or the azimuthal force. I understand this force is something you introduce unnecessarily? Also, some explanation why uperp2 does not come from the perp equation would be good}
%\eag{What happens with the third term?}
%In the collisionless case, $u_{\perp e0}=0$, so the term $n_0 u_{e\perp 0} h_{e2}$ in Eq. \eqref{ql:flux} is negligible.
The normalized values for the two terms constituting the second-order electron flux of Eq. \eqref{ql:flux}, $n_0 u_{\perp e2}$ and $n_0 \Braket{ u_{\perp e1} h_{e1}^*}$, have been plotted in Figure \ref{fig:ql:u2} in the case of the simulation data from [\onlinecite{jime23a}].
From the figure it can be seen that the second order cross-field velocity is directed inwards in those regions where $\nabla_\perp \ln (n_0/B^2) < 0$ and outwards when the gradient changes sign, while the flux contribution due to the $n_0 \Braket{ u_{\perp e1} h_{e1}^*}$ term presents the opposite behavior.
%\eag{where do I see this?}
It is the competition between these two terms that will
dictate whether quasi-linear transport is inward or outward in the MN.

Substituting the expressions for $\Braket{ u_{\perp e1} h_{e1}^*}$ and $u_{\perp e2}$ in Eq. \eqref{ql:flux} and discarding any $O(\epsilon)$ term,
%:\eag{Is this always true in our 0th order solution?}
%
\begin{align}
    \left( n_e u_{\perp e} \right)_2 
    =
    n_0 c_e k^2 \rho_e^3 \w_{LH}^2
    \frac{k_\theta \w_{ri} \gamma}{\left(\w_{ri}^2 + \gamma^2\right)^2}
    \frac{e^2 |\phi_1|^2}{T_e^2} \ \text{.}
    \label{2nd_el_flux}
\end{align}
%

%(anunciar mejor lo que vas a decir)
%Being $\w_{ri} k_\theta > 0$ for $u_{\theta e0}>0$, as in Eq. \eqref{LF:lhdi:gamma} and in Figure \ref{imag:MN:2D_MN_rate}, 
%\eag{I am not sure this comment should be here}
%the total second order flux $\left( n_e u_{\perp e} \right)_2$ is positive (directed outwards).
%
%(explicar mejor que esto va en dirección de cierto gradiente, si no es un salto mental grande:)
Eq. \eqref{2nd_el_flux} shows that the second order electron flux has the same sign of the second order velocity $\Braket{u_{\perp e1} h_{e1}^*}$. Thus, the cross-field electron current is directed against the perpendicular gradient of $n_0 / B^2$. This, in practice and for radially-decreasing plasma density profiles, means that quasi-linear transport is directed away from the axis.
Quasi-linear diffusion then effectively acts to quench plasma inhomogeneities.
This highlights the role that instabilities have in inhomogeneous plasmas, which is that of attenuating gradient-driven drifts present at the equilibrium. Unstable oscillations, induced by the presence of said drifts, cause the plasma to migrate in a direction opposite to that of the equilibrium gradients, relaxing the inhomogeneities and therefore the induced drifts.

%\subsection{Saturation condition}
 
%----------------------------------------------------------------------------------
%----------------------------------------------------------------------------------
%----------------------------------------------------------------------------------

\section{Summary}

\label{sec:conclusion}

We have derived a local linear stability model of $3$D electrostatic isothermal waves in a partially magnetized plasma presenting inhomogeneities in both parallel and perpendicular directions of the magnetic field, taking into account magnetic curvature effects, parallel dynamics, gyroviscosity, collisionality, and inertial effects from a fluid perspective.
%\mmm{rather, mention gyroviscosity explicitly}
The equilibrium plasma temperature has been assumed to be isotropic.
The inclusion of $2$D gradients of both zeroth and first-order plasma quantities represents a novelty of this work with respect to the available literature.

The presented model is based on expansions of the momentum and continuity equations in the small parameter $\epsilon = {\rho_e}/{L}$. We have shown the detailed derivation of the dispersion relation in the Low Frequency regime, that is, $\w_e = O(\w_{ce} \ \epsilon)$. Our proposed approach is particularly convenient for assessing the effect of plasma inhomogeneities by iteratively including larger powers of $\epsilon$ as shown in an earlier version of this work, presented as a conference paper in [\onlinecite{ripo24a}], detailing the application of our approach to the High Frequency regime. 
%\mmm{one sentence saying that our matrix approach is novel (and convenient)}

From this analysis, we have provided simple and general instability criteria for both drift-gradient and drift-dissipative instabilities, highlighting the role of each drift and their interplay on the onset of drift-driven unstable oscillations and the effect of finite Larmor radius effects on the onset/quenching of exponential growth.

The dispersion relation has been specialized to the study of oscillations in the MN of a Helicon Thruster, using the data from [\onlinecite{jime23a}] as a reference. Maps of the growth rate, real oscillation frequency and wavenumber have been provided for the most unstable modes predicted to arise in the MN, showing the onset of essentially-azimuthal instabilities in the $1$ kHz--$1$ MHz range.
The analysis has highlighted the importance of including parallel inhomogeneities in the formulation of the dispersion of an $E\times B$ plasma discharge such as that of a MN, as these gradients may drive instabilities even when conditions for Modified Simon-Hoh Instability are not met and in the absence of axial propagation.
Three regions of the MN can be identified: one in the upper near-plume, where perpendicular gradients are dominant, and where conditions for MSHI are respected (point A of Figure \ref{imag:pedro}); one in the lower near-plume, where conditions for MSHI are not respected, and where both perpendicular and parallel dynamics effects become important in triggering instabilities (point B); one in the far-plume, where far effects are dominant (point C). 
Collisional effects are shown to be secondary, at least in the explored parametric range.
While based on a local, linear approximation, these findings are in qualitative agreement with some of the available experimental data, where significant spectral power density is found in this frequency range in several MNs \cite{taka22a, hepn20b, vinci22e, madd24b} %\mmm{cites}
.
%This qualitative assessment of linear local fluid instabilities in a reference MN case shows that a vast region of the discharge is linearly unstable. The validity of this study requires said oscillations to remain small with respect to equilibrium quantities; to assess which of these instabilities may eventually grow into sizable unstable phenomena, a more complex non-linear stability analysis should be carried out, going beyond the scope of this paper.

As a final step, we have carried out a quasi-linear analysis to gauge the second-order effect of oscillations on transport in the absence of collisions. From this study, relevant to the aforementioned low frequency regime, we have obtained the two main components of the second order electron flux, $n_0 u_{\perp e2}$ and $n_0 \Braket{u_{\perp e1} h_{e1}^*}$. The former is directed as the gradient of $n_0 /B^2$, while the latter is in the opposite direction; being the latter larger, $(n_e u_{\perp e})_2$ is directed against the density gradient.
%
%derived two important conclusions: instabilities induce a cross-field electron velocity $u_{\perp e2}$ 
%in the same direction of $\nabla_\perp \ln (n_0/B)$; at the same time, the net cross-field electron flux $(n_e u_{\perp e})_2$ 
%is directed in the opposite direction, inducing a second order correction on the electron number density causing a relaxation of the plasma gradients.
%\mmm{estresar más la discusión de la dirección}
Then, the role of instabilities seems to be that of `pushing' the plasma against the equilibrium gradients, attenuating the zeroth-order drifts which cause the plasma to destabilize in the first place. This conclusion agrees with the observations of Hepner et al. \cite{hepn20b} of an outward electron flux, in contrast with the description from Takahashi et al. \cite{taka22a} of an inward particle flux.

We stress that for this fluid model to be valid we are intrinsically limited to the study of long wavelengths, $k \rho_e < 1$, and small parallel propagation, $|c_e k_\prl| < |\w_e|$. Moreover, having assumed isothermal perturbations as a closure may not always be valid. To overcome these limitations, a more consistent as well as complex kinetic approach should be employed.

Moreover,
as these instabilities grow in time, their amplitude may become comparable to the zeroth-order plasma quantities, violating the validity of the present linear stability analysis as non-linear effects become important. To understand which of these instabilities may evolve into appreciably large amplitude oscillations a more complex non-linear stability should be carried out. This, however, goes beyond the scope of this paper.

%----------------------------------------------------------------------
\section*{Acknowledgments}
%----------------------------------------------------------------------

This project has received funding from the European Research Council (ERC) under the European Union’s Horizon 2020 research and innovation programme (Starting Grant project ZARATHUSTRA, grant agreement No 950466).

\section*{Data availability}

The simulation dataset used in this work can be accessed freely on \url{https://doi.org/10.5281/zenodo.10641401} %\cite{sim_data_jime23a}
under the license terms detailed therein.

% \section*{Additional material}

% Notebooks

\appendix

%----------------------------------------------------------------------------------
%----------------------------------------------------------------------------------
%----------------------------------------------------------------------------------

\section{Gyroviscous Tensor Divergence}

\label{app:gyro}

From the definition of the gyroviscous tensor, assuming isotropic temperature $T_{\perp e} = T_{\prl e}$ and fast-ordering dynamics \cite{ramo05b}:
\begin{widetext}
\begin{multline}
    - \frac{\nabla \cdot \Pi_e}{m_e n_e} = 
    \frac{1}{e n_e} \Bigg[ \left( \nabla \times \left( \frac{T_e n_e}{B} \bm 1_\prl \right) \right) \cdot \nabla \bm u_e - \frac{\nabla}{2} \left( \frac{T_e n_e}{B} \bm 1_\prl \cdot (\nabla \times \bm u_e) \right) \\
    + B \bm 1_\prl \cdot \nabla \bigg( \frac{T_e n_e}{B^2} \bigg( 3 \bm 1_\prl \times (\bm 1_\prl \cdot \nabla \bm u_e) + \bm 1_\prl \times (\bm 1_\prl \times (\nabla \times \bm u_e))
    + \bm 1_\prl \cdot \frac{\nabla \times \bm u_e}{2} \bm 1_\prl \bigg) \bigg) \\ 
    - \nabla \times \bigg( \frac{T_e n_e}{B} \bigg( \bm 1_\prl \cdot \nabla \bm u_e
    + \frac{\bm 1_\prl}{2} \left( \nabla \cdot \bm u_e - 3 \bm 1_\prl \cdot \nabla \bm u_e \cdot \bm 1_\prl \right) \bigg) \bigg) \Bigg] \ \text{;}
\end{multline}
\end{widetext}
from its expression, it is apparent that at the zeroth order $(\nabla \cdot \Pi_{e} / (m_e n_e))_0 = O( \w_{ce} u_{\theta e0}\ \epsilon^2)$. \\
At the first order,
%assuming isothermal fluctuations $T_{e1}=0$ and
assuming $\bm u_{e0} \simeq u_{e\theta0} \bm 1_\theta$ and neglecting $O(\epsilon^2)$ terms:
\begin{widetext}
    \begin{multline}
        \left( - \frac{\nabla \cdot {\Pi}_e}{m_e n_e} \right)_{\perp1} =
        \frac{c_e^2}{\w_{ce}} \Bigg\{  \Bigg[ \frac{k_\perp^2 + k_\theta^2 + 2 k_\prl^2}{2} 
        - i k_\perp \left( \nabla_\perp \ln {u}_{e\theta1} + \frac{1}{2} \nabla_\perp \ln \left( \frac{p_{e0}}{B^2} \right) \right)
        - 2 i k_\prl \left( \nabla_\prl \ln {u}_{e\theta1} + \frac{1}{2} \nabla_\prl \ln \left( \frac{p_{e0}}{B^3} \right) \right) \Bigg] u_{e\theta1} \\
        -i \frac{i k_\theta}{2} \nabla_\perp \ln \left( \frac{p_{e0}}{B^2} \right) u_{e\perp1}
        - i k_\theta \left[ i k_\prl + \nabla_\prl \ln {u}_{e\prl1} + \frac{1}{2} \nabla_\prl \ln \left( \frac{p_{e0}^2}{B^5} \right) \right] u_{e\prl1} \Bigg\}
        - i \frac{c_e^2}{2 \w_{ce}} \left(k_\perp \nabla_\perp + 2 k_\prl \nabla_\prl \right) u_{e\theta0} \frac{p_{e1}}{p_{e0}}
    \end{multline}
    \begin{multline}
        \left( - \frac{\nabla \cdot {\Pi}_e}{m_e n_e} \right)_{\theta1} =
        \frac{c_e^2}{\w_{ce}} \Bigg\{ - \Bigg[ \frac{k_\perp^2 + k_\theta^2 + 2 k_\prl^2}{2}
        - i k_\perp \left( \nabla_\perp \ln {u}_{e\perp1} + \frac{1}{2} \nabla_\perp \ln \left( \frac{p_{e0}}{B^2} \right) \right)
        - 2 i k_\prl \left( \nabla_\prl \ln {u}_{e\perp1} + \frac{1}{2} \nabla_\prl \ln \left( \frac{p_{e0}}{B} \right) \right) \Bigg] u_{e\perp1}\\
        -i \frac{i k_\theta}{2} \nabla_\perp \ln \left( \frac{p_{e0}}{B^2} \right) u_{e\theta1}
        + i \bigg[ k_\perp \left( i k_\prl + \nabla_\prl \ln {u}_{e\prl1} + \frac{1}{2} \nabla_\prl \ln \left( \frac{p_{e0}^2}{B^5} \right) \right)
        + k_\prl \left( \nabla_\perp \ln u_{e\prl1} + \nabla_\perp \ln B \right) \bigg] u_{e\prl1} \Bigg\} \\
        + i \frac{c_e^2}{2 \w_{ce}} \left( k_\theta \nabla_\perp u_{e\theta0} \right) \frac{p_{e1}}{p_{e0}}
        \label{gyro:theta}
    \end{multline}
    \begin{multline}
        \left( - \frac{\nabla \cdot {\Pi}_e}{m_e n_e} \right)_{\prl1} =
        \frac{i c_e^2}{\w_{ce}} \Bigg\{ k_\theta \bigg[ i k_\prl + \nabla_\prl \ln {u}_{e\perp1}
        + \frac{1}{2} \nabla_\prl \ln B \bigg] u_{e\perp1}
        - \bigg[ k_\perp \left( i k_\prl + \nabla_\prl \ln {u}_{e\theta1}
        + \frac{1}{2} \nabla_\prl \ln B \right) \\
        + k_\prl \left( \nabla_\perp \ln u_{e\theta1} + \nabla_\perp \ln \left( \frac{p_{e0}}{B^3} \right) \right) \bigg] u_{e\theta1} 
        - k_\theta \nabla_\prl \ln \left( \frac{p_{e0}}{B^4} \right) u_{e\prl1} \Bigg\}
        - i \frac{c_e^2}{2 \w_{ce}} \left( k_\perp \nabla_\prl u_{e\theta0} \right) \frac{p_{e1}}{p_{e0}} \ \text{.}
    \end{multline}
    %\newpage
\end{widetext}
At the second order, discarding any term of order $O\left(  \epsilon \right)$ with respect to $\w_{ce} \bm u_{e2}$,
\begin{multline}
    - \frac{\nabla \cdot \Pi_{e2}}{m_e n_0} = \frac{c_e^2}{\w_{ce}} \Bigg[ \left( \bm k \times \bm 1_\prl \right) \cdot \bm k \Braket{i h_{e1} \left( i \bm u_{e1} \right)^*} \\
    - \frac{\bm k}{2} \Braket{ i h_{e1} \left( \bm 1_\prl \cdot i \bm k \times \bm u_{e1}\right)^*} + \frac{\bm k}{2} \Braket{h_{e1}  \bm 1_\prl \cdot \bm k \times \bm u_{e1}^*}  \Bigg] =\\
    \frac{c_e^2}{\w_{ce}} \frac{\bm k}{2} \Bigg[ \bm 1_\prl \cdot \bm k \times 
    \bigg( - \frac{i h_{e1} \left( - i \bm u_{e1}^* \right) - i h_{e1}^*  i \bm u_{e1} }{4} \\
    + \frac{h_{e1} \bm u_{e1}^* + h_{e1}^* \bm u_{e1}}{4} \bigg) \Bigg] \exp(-2\gamma t) = 0
    \ \text{.}
    \label{gyro:2nd}
\end{multline}
Then, $(\nabla \cdot \Pi_{e} / (m_e n_e))_2$ yields no contribution of order $O(\w_{ce} \bm u_{e2})$ in the second order for of the momentum equation \eqref{2nd:el:mom}.

%----------------------------------------------------------------------------------
%----------------------------------------------------------------------------------
%----------------------------------------------------------------------------------

%\begin{comment}

\section{Wave amplitude in presence of zeroth-order parallel gradients}

\label{app:grad}

%As shown in section \ref{subsec:shape}, 
The presence of gradients in our plasma imply that a wave traveling in the same direction of the gradients will experience a change in effective medium impedance as they propagate. 
%\mmm{why is this a problem. Poner aqui eqs B1-B3. Y explicar el problema (la solución ha de respetar esto) y por qué no es satisfactorio para nosotros, y luego ya nuestra solución}
This causes terms in the form $\nabla_\prl Q_0 \nabla_\prl Q_1$ to appear in the expression of the dispersion relation. If we give $Q_1$ an arbitrary shape, we are inadvertently introducing terms acting as artificial sources or sinks or energy.
%Our solution has to keep this change of impedance into consideration, adding a `secular' variation on top of the usual spatial oscillations of the fields.
To show this, we next discuss a wave propagating in the parallel direction, with arc length coordinate $\sigma$, described by the $1$D wave equation for a generic perturbed quantity $ \Psi_1(\w,\sigma)$
\begin{align}
    \frac{\pd^2  \Psi_1}{\pd \sigma^2} = - k^2_\sigma (\w,\sigma) \Psi_1 \ \text{,}
    \label{wave_eq}
\end{align}
with $k_\sigma$ real-valued function of $\w$ and $\sigma$, and with
\begin{align}
    \left|\frac{\pd k_\sigma}{\pd \sigma}\right| \ll k_\sigma^2 \ .
    \label{WKB_approx}
\end{align}

We can use the WKB method to find a solution in the form of
\begin{align}
    \Psi_1 (\w,\sigma) \propto \exp \left( \pm i \int_{\sigma_0}^\sigma k_\sigma (\w,s) ds \right)
    \label{WKB}
\end{align}
which approximately satisfies Eq. \eqref{wave_eq} as long as \eqref{WKB_approx} is valid.
Rewriting the same equation in a new reference frame $\zeta = \zeta(\sigma)$ yields, assuming $\zeta(\sigma)$ to be locally invertible and with $\tilde \Psi_1(\w,\zeta) = \Psi_1 (\w,\sigma)$:
\begin{align}
    & \frac{\pd^2 \tilde \Psi_1}{\pd \zeta^2} + \kappa \left( \zeta \right) \frac{\pd \tilde \Psi_1}{\pd \zeta} = 
    - \left( \frac{\pd \zeta}{\pd \sigma} \right)^{-2} k_\sigma^2 \left(\w, \sigma (\zeta) \right) \tilde \Psi_1 \ \text{,}  \label{eikonal} \\
    & \text{ with} \quad \kappa \left( \zeta \right) = \frac{\pd^2 \zeta}{\pd \sigma^2} \left( \frac{\pd \zeta}{\pd \sigma} \right)^{-2}\ \notag .
    %\text{and} \quad | k_\zeta | = \left| \frac{\pd \sigma}{\pd \zeta} k_\sigma \right| \  \notag .
\end{align}
Here, a solution in the form of $\tilde \Psi_1 \propto \exp(\pm i \int k'(\w,\zeta) d\zeta)$, with $k'$ real-valued function of $\w$, cannot exist, as the left-hand side of Eq. \eqref{eikonal} would include a term $\pm i \kappa(\zeta) k'(\w, \zeta)$, while the right-hand side would still be a real-valued function of $\w$.
Instead, applying the WKB method once again, we find a solution in the form
\begin{align}
    \tilde \Psi_1 (\w,\zeta) \propto \exp \left[ \int_{\zeta_0}^\zeta \left(-\frac{\kappa (s)}{2} \pm i k_\zeta (\w,s) \right) ds \right]
    \label{WKB_zeta}
\end{align}
with
\begin{align}
    k_\zeta^2 (\w,\zeta) = \left( \frac{\pd \zeta}{\pd \sigma} \right)^{-2} k_\sigma^2 \left(\w, \sigma (\zeta) \right) - \frac{\kappa^2(\zeta)}{4}
    \label{k_zeta}
\end{align}
a real-valued function of $\w$. The solution in \eqref{WKB_zeta}, valid as long as the inequality in Eq. \eqref{WKB_approx} is applicable to both $\kappa$ and $k_\zeta$, substituting $\sigma$ with $\zeta$, does not incur in the same problem discussed before, as both sides of Eq. \eqref{eikonal} would be real-valued functions of $\w$. 
This formulation, therefore, does not induce an artificial instability onto waves in the $\zeta$ reference system.

Our dispersion relation is formally the same as that expressed in Eq. \eqref{eikonal}. Indeed, defining $\Phi_1 = n_0 \phi_1 B^{-1}$, the equation
\begin{multline}
    \nabla_\prl^2 [\Phi_1 \exp(i k_\prl s_\prl)] - \nabla_\prl \ln \frac{p_{e0}}{B} \nabla_\prl [\Phi_1 \exp(i k_\prl s_\prl)] = \\
    - \frac{\w_\prl + i \nu_e}{c_e^2} \left[ \w_{Me} - \frac{k^2 c_s^2 (\w_{Te} + \w_e)}{\w_i^2 - k^2 c_s^2} \right] \Phi_1 \label{LF:eikonal}
\end{multline}
corresponds to the $k \rho_e \to 0$ limit of the dispersion relation from Eq. \eqref{LF:disp_rel}, with the definition of $\W_\prl^2$ from Eq. \eqref{LF:R_prl:2} and assuming the inequalities
\begin{align*}
    \left| \nabla_\prl k_\prl \right| \ll k_\prl^2 \ , \qquad \left| \nabla_\prl^2 \ln \Phi_1 \right| \ll \left( \nabla_\prl \ln \Phi_1 \right)^2
\end{align*}
to be valid.
By applying the change of coordinates $\xi = \xi(s_\prl)$, with
\begin{align}
    \nabla_\prl \ln \nabla_\prl \xi = \nabla_\prl \ln \frac{p_{e0}}{B} \ ,
\end{align}
we get, with $\tilde \Phi_1(\w,\xi) = \Phi_1 (\w,s_\prl) \exp(i k_\prl s\prl)$,
\begin{multline}
     \frac{\pd^2 \tilde \Phi_1}{\pd \xi^2} = - \frac{\w_\prl + i \nu_e}{c_e^2 (\nabla_\prl \xi)^2} \left[ \w_{Me} - \frac{k^2 c_s^2 (\w_{Te} + \w_e)}{\w_i^2 - k^2 c_s^2} \right] \tilde \Phi_1\\
    \equiv - k^2_\xi (\w,\xi) \tilde \Phi_1 \ ,
    \label{wave_eq:zeta}
\end{multline}
which is exactly the same as Eq. \eqref{wave_eq}. Since $k_\xi^2$ is a real-valued function of $\w$ for $\nu_e=0$, it follows that $k_\prl^2$ has to be real-valued as well, as in Eq. \eqref{k_zeta}: subsequently, from Eq. \eqref{WKB_zeta}, the shape of $\Phi_1$ has to be
\begin{align}
    \nabla_\prl \ln \frac{n_0 \phi_1}{B} = \frac{1}{2} \nabla_\prl \ln \frac{p_{e0}}{B} \ ,
\end{align}
which is the $k \rho_e \to 0$ limit of Eq. \eqref{LF:phi_shape}.
The same result can be reached by following the procedure shown in section \ref{sec:disp_rel_deriv}, where we imposed the imaginary part of $\W_\prl^2$ to be zero. The two approaches can be summarized as follows: parallel gradients don't introduce complex coefficients in the differential wave equation. Collisions, on the other hand, are energy sinks, therefore they introduce an imaginary coefficient in the dispersion relation as shown in equations \eqref{wave_eq:zeta} and \eqref{LF:disp_rel}, forcing $\w$ to be complex \cite{STIX92}. A similar procedure should be applied to determine the shape of $\nabla_\perp \ln \phi_1$ if we were to extend the analysis to the $O(\epsilon^2)$ order.
%; in other words, if $k_x(\w)$ describes a wave in a conservative medium, equation \eqref{eikonal:eq} (which explicitly includes gradients) must describe the same wave.

%\end{comment}

% \textrm{Re} \left\{  \right\} \cdot \textrm{Re} \left\{  \right\}

%----------------------------------------------------------------------------------
%----------------------------------------------------------------------------------
%----------------------------------------------------------------------------------

%----------------------------------------------------------------------------------
%----------------------------------------------------------------------------------
%----------------------------------------------------------------------------------

\bibliographystyle{ieeetr}
\bibliography{bibtex/ep2,bibtex/others,bibtex/litreview}

\end{document}